\documentclass[twocolumn,superscriptaddress,showpacs,pre,amsmath,amssymb]{revtex4}
\usepackage{epsfig}
\begin{document}

\title{Cross-Correlation of Instantaneous Phase Increments in Pressure-Flow Fluctuations:
Applications to Cerebral Autoregulation}

\author{Zhi~Chen}
\affiliation{Center for Polymer Studies and Department of Physics,
               Boston University, Boston, Massachusetts 02215, USA\\}
\author{Kun~Hu}
\affiliation{Center for Polymer Studies and Department of Physics,
               Boston University, Boston, Massachusetts 02215, USA\\}
\affiliation{Division of Gerontology, Harvard Medical School, Beth
Israel Deaconess Medical Center, Boston, Massachusetts 02215, USA\\}

\author{H.~Eugene~Stanley}
\affiliation{Center for Polymer Studies and Department of Physics,
               Boston University, Boston, Massachusetts 02215, USA\\}
\author{Vera~Novak}
\altaffiliation{VN and PChI contributed equally to this work. VN
designed the clinical study, provided data and guidance of the
physiological aspects.  PChI proposed the instantaneous phase
cross-correlation method, and guided the computational aspects.}
\affiliation{Division of Gerontology, Harvard Medical School, Beth
Israel Deaconess Medical Center, Boston, Massachusetts 02215, USA\\}

\author{Plamen~Ch.~Ivanov}
\altaffiliation{VN and PChI contributed equally to this work. VN
designed the clinical study, provided data and guidance of the
physiological aspects.  PChI proposed the instantaneous phase
cross-correlation method, and guided the computational aspects.}
\affiliation{Center for Polymer Studies and Department of Physics,
               Boston University, Boston, Massachusetts 02215, USA\\}
\date{\today}
\begin{abstract}

We investigate the relationship between the blood flow velocities
(BFV) in the middle cerebral arteries and beat-to-beat blood
pressure (BP) recorded from a finger in healthy and post-stroke
subjects during the quasi-steady state after perturbation for four
different physiologic conditions: supine rest, head-up tilt,
hyperventilation and CO$_2$ rebreathing in upright position. To
evaluate whether instantaneous BP changes in the steady state are
coupled with instantaneous changes in the BFV, we compare dynamical
patterns in the instantaneous phases of these signals, obtained from
the Hilbert transform, as a function of time. We find that in
post-stroke subjects the instantaneous phase increments of BP and
BFV exhibit well pronounced patterns that remain stable in time for
all four physiologic conditions, while in healthy subjects these
patterns are different, less pronounced and more variable. We
propose a new approach based on the cross-correlation of the
instantaneous phase increments to quantify the coupling between BP
and BFV signals. We find that the maximum correlation strength is
different for the two groups and for the different conditions. For
healthy subjects the amplitude of the cross-correlation between the
instantaneous phase increments of BP and BFV is small and attenuates
within 3-5 heartbeats. In contrast, for post-stroke subjects, this
amplitude is significantly larger and cross-correlations persist up
to 20 heartbeats. Further, we show that the instantaneous phase
increments of BP and BFV are cross-correlated even within a single
heartbeat cycle. We compare the results of our approach with three
complementary methods: direct BP-BFV cross-correlation, transfer
function analysis and phase synchronization analysis. Our findings
provide new insight into the mechanism of cerebral vascular control
in healthy subjects, suggesting that this control mechanism may
involve rapid adjustments (within a heartbeat) of the cerebral
vessels, so that BFV remains steady in response to changes in
peripheral BP.


\end{abstract}
\maketitle

\section{Introduction} \label{secintr}

Cerebral autoregulation (CA) is the ability of cerebral blood
vessels to maintain steady cerebral perfusion in response to
fluctuations of systemic blood pressure (BP), postural changes or
metabolic demands. This regulatory mechanism is known to operate
over a range of blood pressure values (e.g. 80 - 150 mm
Hg)~\cite{Lassen59} and on time scales of a few
heartbeats~\cite{Narayanan01}. The long-term CA compensates for
chronic BP elevations and metabolic demands~\cite{Panerai98}.
Ischemic stroke is associated with an impairment of
autoregulation~\cite{Schwarz02,Eames02}, that may permanently affect
cerebrovascular reactivity to chemical and blood pressure
stimuli~\cite{Novak03,Novak04}. With impaired CA, the cerebral blood
flow follows BP fluctuations, posing a risk of insufficient blood
supply to the brain during transient reductions in peripheral BP.
Therefore, evaluating the effectiveness of CA is of great interest,
given the clinical implications.

Traditional experimental methods evaluating the mechanism of CA
require time-consuming invasive procedures~\cite{Kety48,Obrist75}
and are focused on long term BP and blood flow velocities (BFV)
characteristics such as the mean, lacking descriptors of temporal
BP-BFV relationship.  To address this problem, an alternative
``dynamic'' approach has been proposed~\cite{Aaslid89} to quantify
CA using transcranial Doppler ultrasound during the transient
responses in cerebral BFV to the rapid BP changes induced
experimentally by thigh cuff inflation, Valsalva maneuver, tilt-up,
or a change in posture~\cite{Panerai98,Tiecks95}. The autoregulation
indices derived from this approach may be more sensitive to
indicators of hypoperfusion after stroke~\cite{Dawson00}.

The analytic methods evaluating the dynamics of cerebral
autoregulation are currently based on mathematical modeling and
Fourier transform analysis~\cite{Panerai99}. The Fourier transform
based transfer function method has been widely
used~\cite{Narayanan01}. This method estimates the relative
cross-spectrum between BP and BFV signals in the frequency domain.
Dynamic indices of autoregulation, based on Fourier transform
methods, presume (i) signal stationarity (i.e., the mean and
standard deviation of the signal are stable and remain invariant
under a time shift) and (ii) a linear BP-BFV relationship. However,
physiologic signals are often nonstationary reflecting transient
responses to physiologic stimuli~\cite{Panerai01}. The effect of
this nonstationarity on the results obtained from the transfer
function analysis has not been carefully assessed in previous
studies.

Here we investigate the dynamics of the BP-BFV relationship when the
system reaches the quasi-steady state after an initial perturbation.
While studies traditionally have focused on the response in BFV to
transient changes in BP~\cite{Panerai98}, we hypothesize that
spontaneous physiologic fluctuations during the quasi-steady state,
which is characterized by absence of physiologic stimuli or constant
level of stimulation, may also contain important information about
the CA mechanism.

Our focus on fluctuations in the BP and BFV is motivated by previous
work which has demonstrated that physiologic fluctuations contain
important information about the underlying mechanisms of physiologic
control. Robust temporal organization was reported for the
fluctuations characterizing cardiac dynamics (inter-beat
intervals)~\cite{Plamennature96,Havlin_phasicaA99,Plamenchaos01,AryPNAS01,Bernaola-Galvan01,Meyer03,Plamen_PhysicaA04,Pavlov05},
respiratory dynamics (inter-breath
intervals)~\cite{suki94,Alencar01,suki03,Mutch05,West05}, locomotion
(gait, fore-arm
motion)~\cite{Hausdorff96,Hausdorff01,Ashkenazy02,Scafetta03,HuPhysicaA04,Ivanov04},
and brain dynamics~\cite{West03,Song05,Bachmann05}. Moreover, it has
been demonstrated that physiologic fluctuations carry information
reflecting the coupling between different physiologic systems, e.g.,
correlations in the heartbeat change with physical
activity~\cite{Karasik02,Martinis03}, with wake and
sleep~\cite{Ivanov99}, during different sleep
stages~\cite{Bunde00,Kantelhardt02,Dudkowska04,Staudacher05}, and
even different circadian phases~\cite{HuPNAS04}. BP and BFV signals
are impacted by the heartbeat, and thus one can expect that
fluctuations in BP and BFV may reflect modulation in the underlying
mechanisms of control. Previous studies have focused on the
transitional changes in BP and BFV in response to abrupt
perturbation of the physiologic state, e.g., rapid switch from
supine to tilt. In contrast, we focus on the dynamical
characteristics of BP and BFV signals after the initial
perturbation, when the system has reached a quasi-steady state
during which there is no change in physiologic stimuli. Further, we
hypothesize that certain dynamical characteristics of the
fluctuations in BP and BFV at the steady state, and how these
characteristics change for different physiologic conditions, may
reflect aspects of the underlying mechanism of CA. For example,
under normal CA the fluctuations of the BFV in healthy subjects at
the steady state may relate to high frequency adjustments (even
within a single heartbeat) of the diameter of the cerebral blood
vessels, while loss of CA after stroke may lead to impaired vascular
dilation or contraction associated with reduced fluctuations in BFV.

\begin{table*}
{
\scriptsize
\begin{tabular}{ccccccccccc}
\hline Variable & & & Demographic & characteristics & & & & & &
\\\hline
&&&&Control&&&Stroke&&&\\

Men/Women&&&&4/7&&&7/6&&&\\

Age(mean$\pm$SD)&&&&$48.2\pm8.7$&&&$52.8\pm7.1$&&&\\

Race W/AA&&&&10/1&&&12/1&&&\\

Stroke side&&&&--&&&5/8&&&\\

(Right/Left)&&&&&&&&&&\\\hline

Group&\multicolumn{2}{c}{Supine}  & \multicolumn{2}{c}{Tilt} &
\multicolumn{2}{c}{Hyperventilation} &\multicolumn{2}{c}{CO$_2$
rebreathing} & \multicolumn{2}{c}{Statistics}\\\cline{2-9}

(mean$\pm$SD)& Control & Stroke & Control & Stroke &Control & Stroke
&Control & Stroke & \multicolumn{2}{c}{(P values)}\\\hline

BP (mm Hg) & $96.4\pm20.9$ & $101.0\pm20.6$ & $93.1\pm20.1$ &
$105.9\pm22.5$ & $94.2\pm19.7$& $105.4\pm22.0$& $97.6\pm21.7$&
$108.8\pm22.7$& 0.66$^{*}$
& 0.0005$^{\dagger}$\\

BFV-MCAR/  & $66.0\pm18.7$ & $55.2\pm18.0$ & $57.3\pm18.1$ &
$49.7\pm18.5$ & $40.3\pm15.7$& $39.8\pm14.9$& $54.7\pm21.9$&
$56.3\pm21.5$& $<0.0001^{*}$
& 0.77$^{\dagger}$\\
Normal side(cm/s) &&&&&&&&&&\\

BFV-MCAL/& $63.5\pm19.6$ & $51.1\pm19.0$ & $54.8\pm17.1$ &
$51.5\pm19.3$ & $40.8\pm14.9$& $41.0\pm17.9$& $54.6\pm21.5$&
$57.4\pm23.3$& $<0.0001^{*}$
& 0.008$^{\dagger}$\\
Stroke side(cm/s) &&&&&&&&&&\\

CO$_2$(mm Hg) & $33.5\pm6.0$ & $37.7\pm4.9$ & $32.0\pm3.6$ &
$32.5\pm2.5$ & $21.0\pm4.5$& $23.5\pm8.0$& $34.6\pm7.2$&
$33.2\pm6.5$&$<0.0001^{*}$
& 0.28$^{\dagger}$\\

CVR**-MCAR  & $1.54\pm0.45$ & $1.96\pm0.54$ & $1.75\pm0.53$ &
$2.28\pm0.93$ & $2.68\pm1.10$& $3.00\pm1.28$& $1.97\pm0.61$&
$2.17\pm0.83$& $0.0001^{*}$
& 0.007$^{\dagger}$\\
/Normal side&&&&&&&&&&\\

CVR-MCAL & $1.59\pm0.56$ & $1.97\pm0.65$ & $1.79\pm0.61$ &
$2.29\pm0.83$ & $2.57\pm0.90$& $3.04\pm1.60$& $1.96\pm0.56$&
$2.19\pm0.94$ & $0.0001^{*}$
& 0.01$^{\dagger}$\\
/Stroke side&&&&&&&&&&\\\hline
\end{tabular}
}

* P value between physiologic conditions comparisons

$\dagger$ P value between groups comparisons

** CVR (cerebral vascular resistance) is defined as mean BP/BFV

\caption{ Demographic characteristics. } \label{table1}
\end{table*}

To test this hypothesis, we measure BP and BFV signals from healthy
and post-stroke subjects during four physiologic conditions: supine,
tilt, hyperventilation, and CO$_2$ rebreathing in upright position.
We apply several complementary methods to quantify the dynamical
BP-BFV relationship in these quasi-steady conditions: transfer
function analysis, cross-correlation and phase synchronization
analyses, and we compare these methods with a {\it new approach} of
cross-correlation between the instantaneous phase increments of BP
and BFV signals. Interactions between peripheral circulation
(beat-to-beat BP) and cerebral vasoregulation [BFV in the middle
cerebral artery (MCA)] can be modeled as dynamic synchronization of
two coupled nonlinear systems. Specifically, We hypothesize that the
CA mechanism may also involve adjustments in the cerebral vascular
tone to spontaneous changes in BP that may be present within a
single heartbeat even when the system is in the steady state and
there are no significant changes in the mean blood pressure.

The synchronization phenomenon was first observed by Hugenii for two
coupled pendulum clocks~\cite{Hugenii73}, and since then it has been
found in many physical and biological systems where two or more
coupled subsystems
interact~\cite{Fabiny93,Anischenko92,Heagy94,Schreiber82,Bahar02,Rybski03,Bahar03}.
Alternatively, the synchronization may also be triggered by the
influence of external noisy or regular
fields~\cite{Pikovsky84,Kuznetsov85}. In recent years, the concept
of synchronization has been widely used to study the coupling of
oscillating systems, leading to the discovery of phase
synchronization in non-identical coupled systems in which the
instantaneous phases are synchronized, while their instantaneous
amplitudes remain
uncorrelated~\cite{Rosenblum96,Rosenblum97,Parlitz96,Bahar04}. Such
phase synchronization has been empirically discovered in a range of
physical and physiological systems~\cite{book_Rosenblum,Osipov}.
Specifically, studies have found coupling between the cardiac rhythm
and other systems: phase synchronization was observed between the
heartbeat and respiration during normal conditions~\cite{Schafer98}
and during respiratory sinus arrhythmia~\cite{Kotani00,Lotric00},
change of cardiorespiratory phase synchronization with new-borns'
age~\cite{Mrowka00} and with the level of
anaesthesia~\cite{StefanovskaPRL00}. Phase analysis methods have
been used to probe spatial synchronization of oscillations in blood
distribution systems~\cite{StefanovskaTPHYS00}, between cortical
centers during migraine~\cite{Angelini04}, as well as between
certain brain areas and muscle activity of the limbs~\cite{Tass98}.

In this study we evaluate the time-domain characteristics of both the
amplitudes and the instantaneous phases of the BP and BFV which can be
considered as two interacting subsystems within the CA mechanism. To
determine the characteristics of the coupling between BP and BFV in
healthy subjects and how they change with stroke, we analyze the
cross-correlation between the instantaneous phase increments of these
two signals. We find that this cross-correlation is much stronger for
post-stroke subjects, indicating increased synchronization between BP and
BFV, which suggests impaired mechanism of the CA. We compare the
results of the instantaneous phase increment cross-correlation
analysis with those obtained from several complementary methods
including the transfer function, cross-correlation and phase
synchronization analyses.

\section{Methods}\label{secmeth}

\subsection{Study Groups}
\medskip

We obtain data from the Autonomic Nervous System Laboratory at the
Department of Neurology at The Ohio State University and from the
SAFE (Syncope and Falls in the Elderly) Laboratory at the Beth
Israel Deaconess Medical Center at Harvard Medical School. All
subjects have signed informed consent, approved by the Institutional
Review Boards.
Demographic characteristics are summarized in Table~\ref{table1}.
Control group: 11 healthy subjects (age $48.2\pm8.7$ years). Stroke
group: 13 subjects with a first minor ischemic stroke ( $>2$ months
after acute onset) (age $52.8\pm7.1$ years). Post-stroke subjects
have a documented infarct affecting $<1/3$ of the vascular territory
as determined by MRI or CT with a minor neurological deficit
(Modified Rankin Score scale $<3$). The side of the lesion is
determined by neurological evaluation and confirmed with MRI and CT.
The lesion is in the right hemisphere in 5 of the subjects and in
the left hemisphere in 8 of the subjects. Normal carotid Doppler
ultrasound study is required for participation. Patients with
hemorrhagic strokes, clinically important cardiac disease including
major arrhythmias, diabetes and any other systemic illness are
excluded. All subjects are carefully screened with a medical
history, physical and laboratory examination.


\begin{figure*}
\centerline{
\epsfysize=0.9\columnwidth{\rotatebox{-90}{\epsfbox{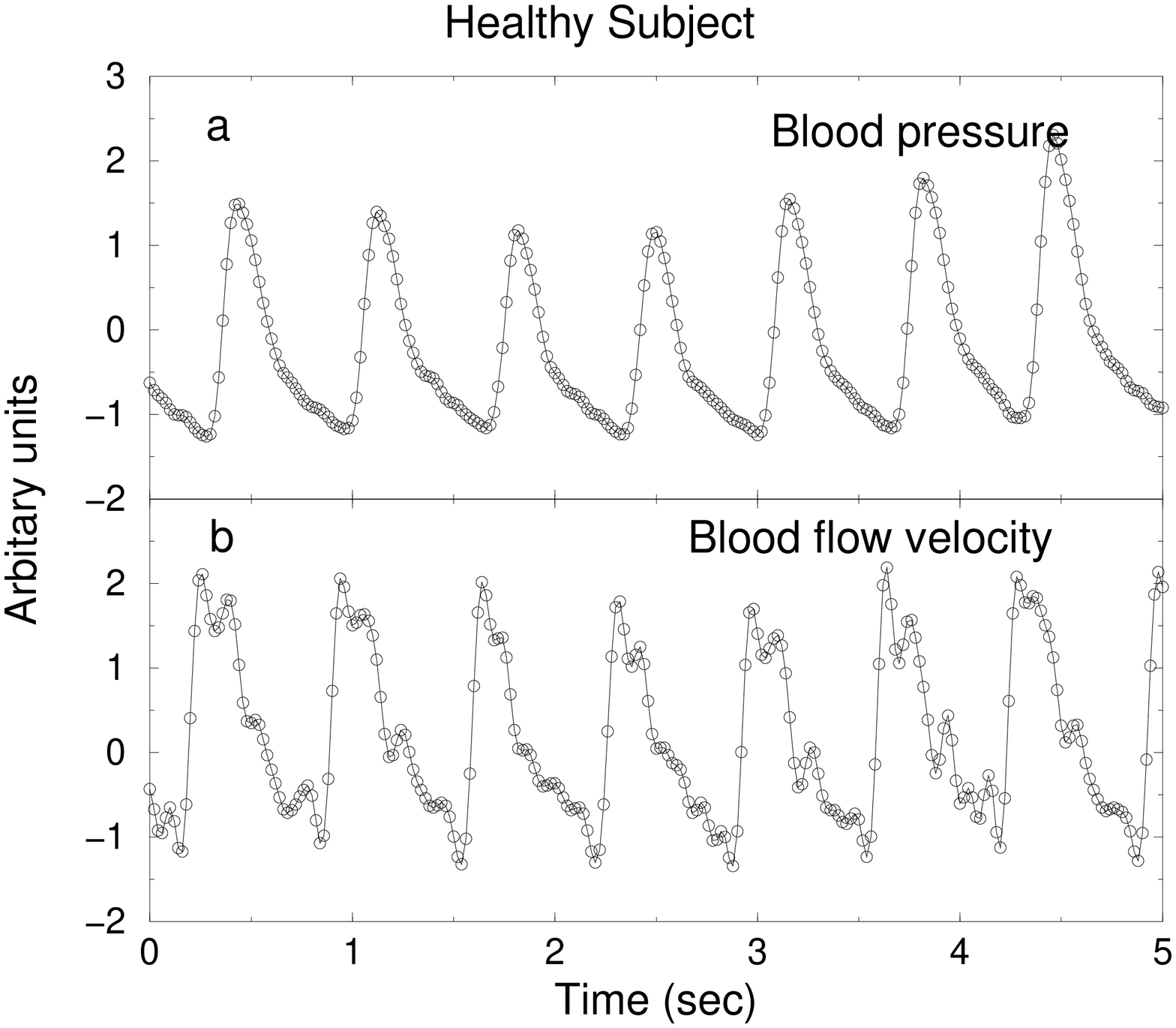}}}
\hspace{0.3cm}
\epsfysize=0.9\columnwidth{\rotatebox{-90}{\epsfbox{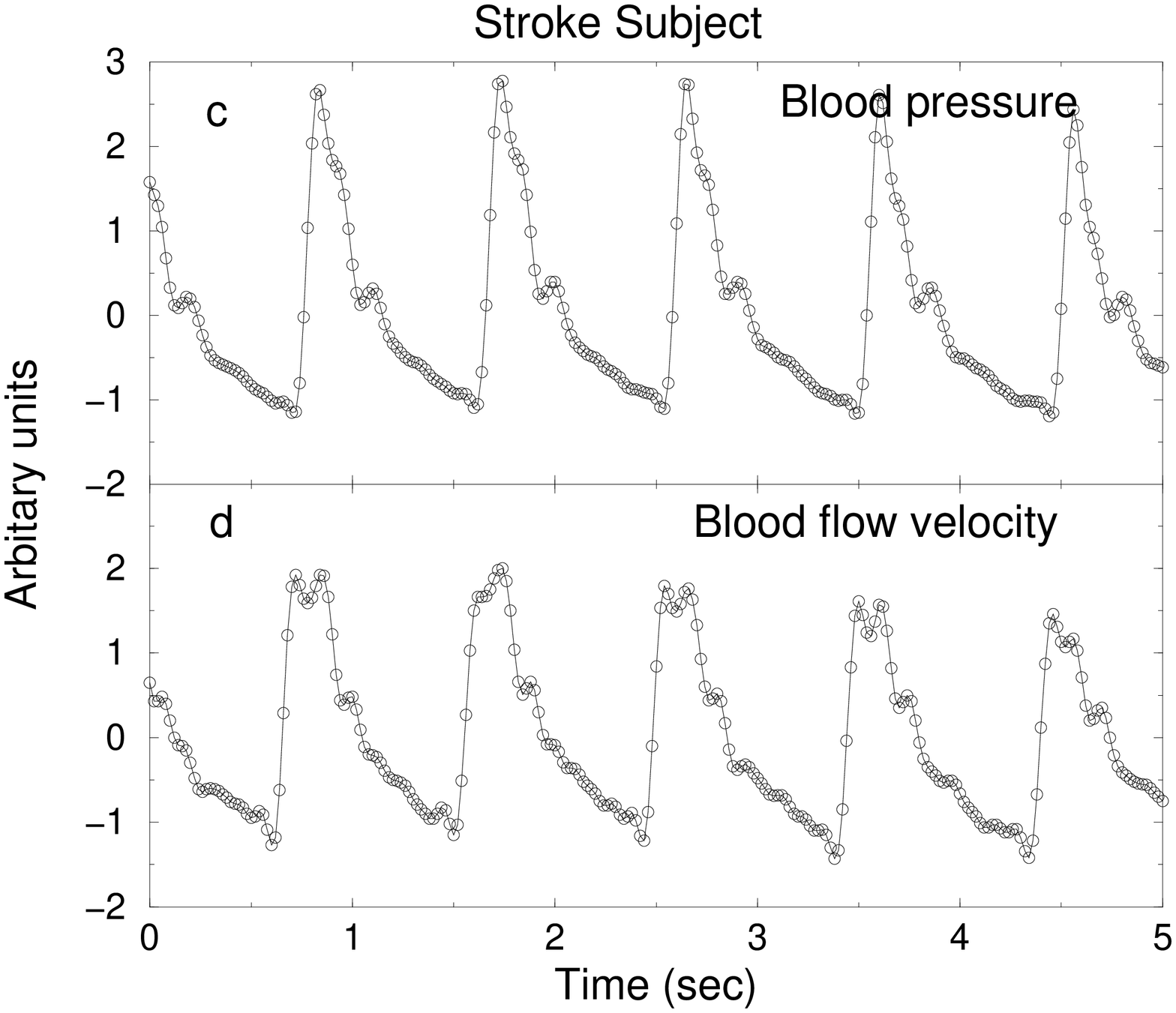}}}}

\caption{BP and BFV signals during CO$_2$ re-breathing after a
band-pass Fourier filter in the range [0.05Hz, 10Hz] and
normalization to unit standard deviation: (a-b) for a healthy
subject, (c-d) for a post-stroke subject.} \label{fig01}
\end{figure*}

\subsection{Experimental Protocol}


All subjects have participated in the following experimental
protocol:

\noindent $\bullet$ Baseline supine rest --- normal breathing
(normocapnia): subject rests in supine position for 5 minutes on a
tilt table;

\noindent $\bullet$ Head-up tilt --- upright normocapnia: The tilt table
is moved upright to an 80$^{\circ}$ angle. The subject remains in
upright position for 5 minutes and is breathing spontaneously;

\noindent $\bullet$ Hyperventilation --- upright hypocapnia: the subject
is asked to breathe rapidly at $\approx$1 Hz frequency for 3 minutes
in an upright position. Hyperventilation induces hypocapnia (reduced
carbon dioxide), which is associated with cerebral vasoconstriction;

\noindent $\bullet$ CO$_2$ rebreathing --- upright hypercapnia: The
subject is asked to breath a mixture of air and $5\%$ CO$_2$ from
rebreathing circuit at a comfortable frequency for 3 minutes in an
upright position. CO$_2$ rebreathing increases carbon dioxide above
normal levels and induces hypercapnia, which is associated with
vasodilatation.

The mechanism of CA is at least partially related to the coupling
between metabolic demands and oxygen supply to the
brain~\cite{Panerai98}. Carbon dioxide (CO$_2$) is one of the most
potent chemical regulators of cerebral vasoreactivity.  Head-up tilt
provides both pressure and chemical stimulus --- BFV and CO$_2$
decline in upright position, reflecting the change in intracranial
pressure and shifting autoregulatory curve towards lower BP values.
There is a linear relationship between CO$_2$ values and cerebral
blood flow: hypocapnia (through hyperventilation) causes
vasoconstriction and thus decreases the blood flow, hypercapnia
(through CO$_2$ rebreathing) causes vasodilatation and increases the
blood flow in the brain~\cite{Panerai98}.

\subsection{Data Acquisition}

We perform experiments in the morning or more than 2 hours after the
last meal. We measure the electrocardiogram from a modified standard
lead II or III using a SpaceLab Monitor (SpaceLab Medical Inc.,
Issaquah, WA). We record beat-to-beat BP from a finger with a
Finapres device (Ohmeda Monitoring Systems, Englewood CO), which is
based on a photoplethysmographic volume clamp method. During the
study protocol, we verify BP by arterial tonometry. With finger
position at the heart level and temperature kept constant, the
Finapres device can reliably track intraarterial BP changes over
prolonged periods of time. We measure the respiratory waveforms with
a nasal thermistor. We measure CO$_2$ from a mask using an infrared
end tidal volume CO$_2$ monitor (Datex Ohmeda, Madison WI). We
insonate the right and left MCAs from the temporal windows, by
placing the 2-MHz probe in the temporal area above the zygomatic
arch using a transcranial Doppler ultrasonography system (MultiDop
X4, DWL Neuroscan Inc, Sterling, VA). Each probe is positioned to
record the maximal BFV and is fixed at a desired angle using a
three-dimensional positioning system attached to the light-metal
probe holder. Special attention is given to stabilizing the probes,
since their steady position is crucial for reliable, continuous BFV
recordings. BFV and all cardiovascular analog signals are
continuously acquired on a beat-to-beat basis and stored for
off-line post-processing. We visually inspect the data and remove
occasional extrasystoles and outlier data points using linear
interpolation. We use the Fourier transform of the Doppler shift
(the difference between the frequency of the emitted signal and the
echo frequency of the reflected signal) to calculate BFV. BFVs in
the MCA correlate with invasive measurements of blood flow with
xenon clearance,  laser Doppler flux and positron emission
tomography~\cite{Nobili96,Leftheriotis95,Sugimori95}. Since the MCA
diameter is relatively constant under physiological
conditions~\cite{Serrador00}, BFV can be used for blood flow
estimates.

\begin{figure*}
\centerline{
\epsfysize=0.9\columnwidth{\rotatebox{-90}{\epsfbox{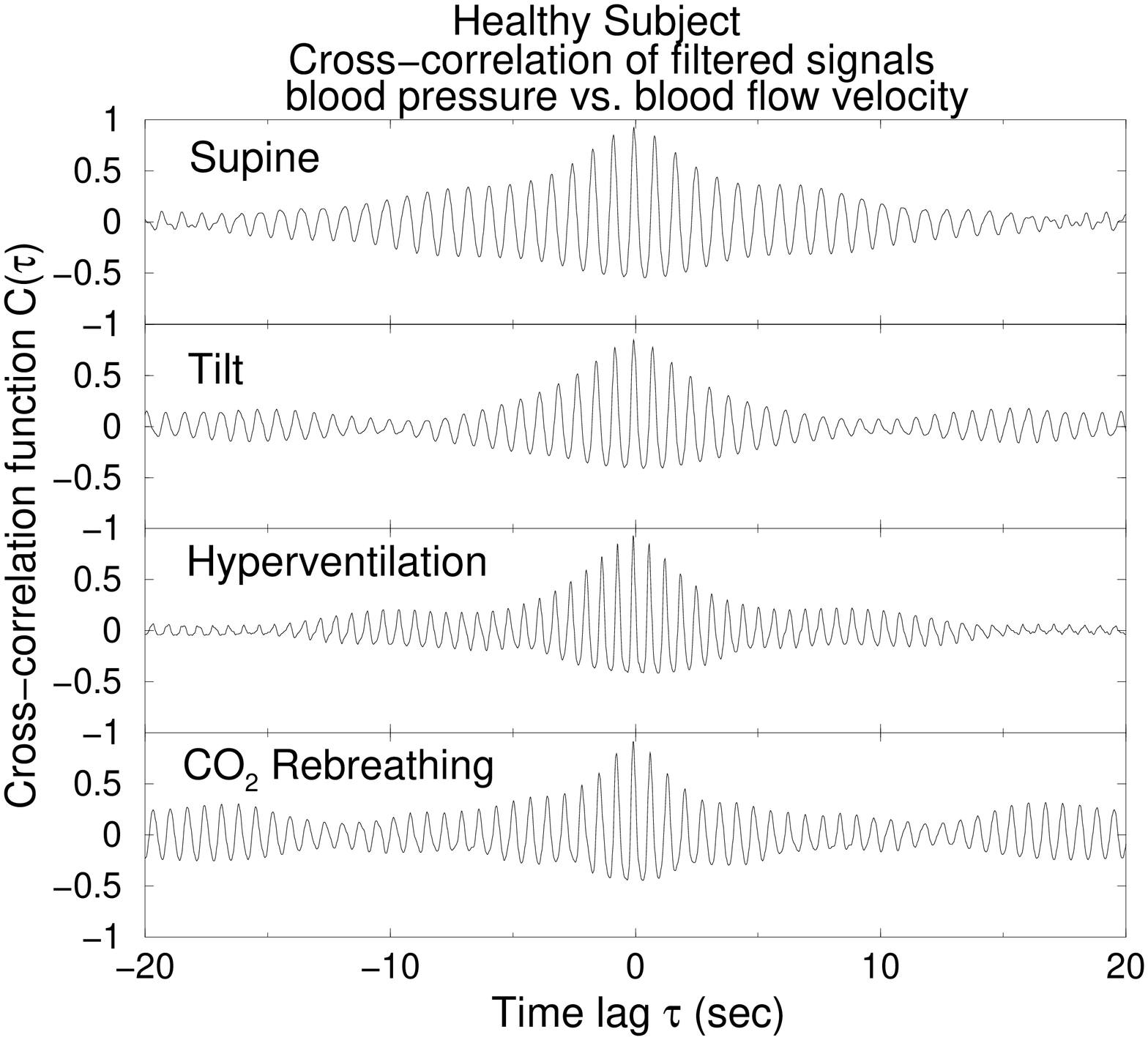}}}
\hspace{0.3cm}
\epsfysize=0.9\columnwidth{\rotatebox{-90}{\epsfbox{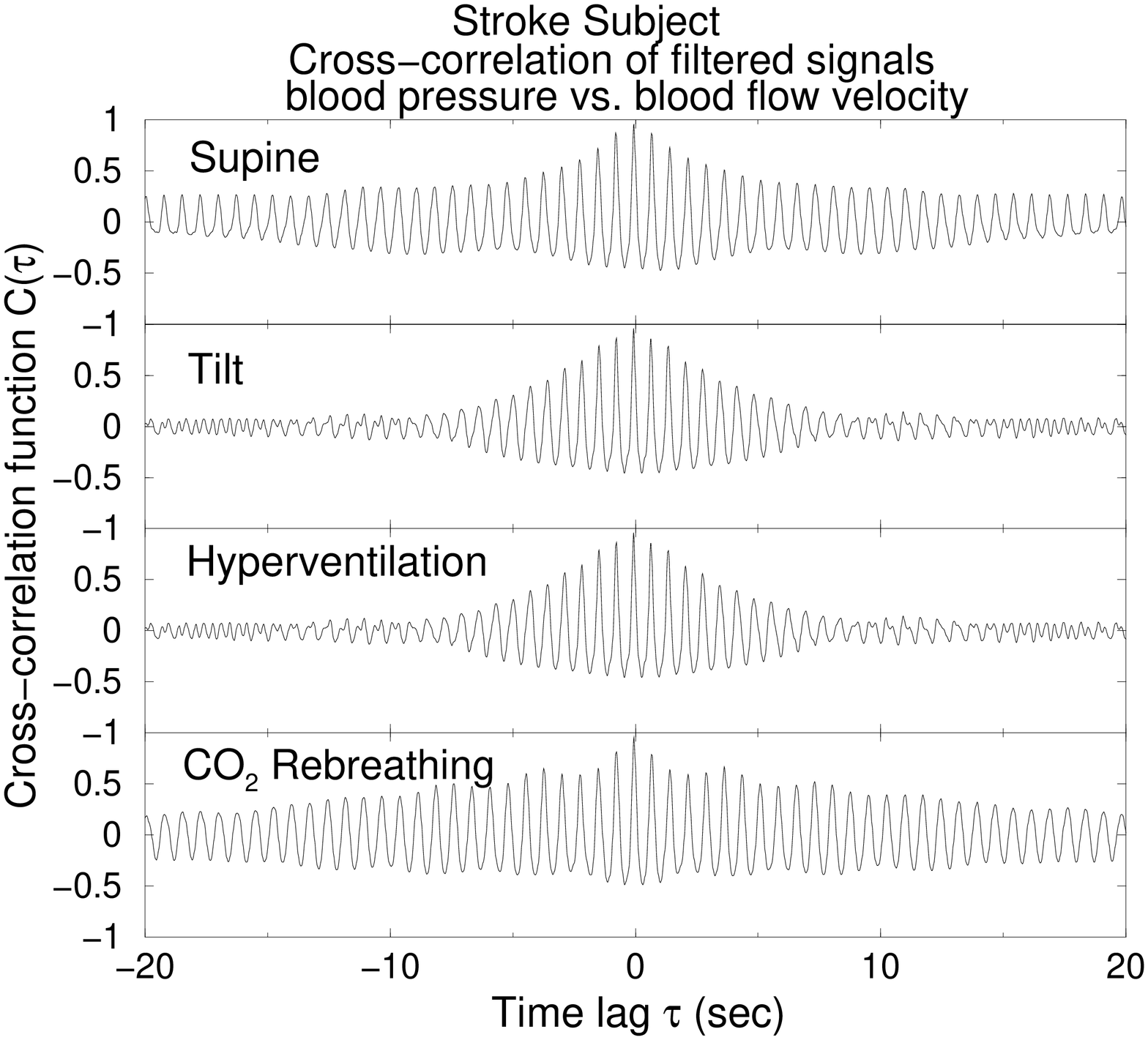}}}}

\caption{Cross-correlation function $C(\tau)$ for the BP and BFV
signals during four physiologic conditions: (a) for the same healthy
subject and (b) for the same post-stroke subject as shown in
Fig.~\ref{fig01}. BP and BFV signals are pre-processed using a
band-pass Fourier filter in the range [0.05Hz, 10Hz] and are
normalized to unit standard deviation before the analysis. Since BFV
precedes BP, the maximum value $C_{\rm max}$ in the
cross-correlation function $C(\tau)$ is located not at zero lag but
at $\tau\approx -0.1 $ sec.} \label{fig01n}
\end{figure*}

\subsection{Statistical Analyses}


We use the multiple analysis of variance (MANOVA) with 2x4 design
for the two groups (control and stroke) and for four physiologic
conditions (supine, tilt, hyperventilation, and CO$_2$ rebreathing)
with subjects as nested random effects (JMP version 5 Software
Analysis Package, SAS Institute, Cary, NC). For each group and
condition, we calculate: (i) the mean BP, BFV, cerebral vascular
resistance (CVR, calculated by mean BP/BFV) from the right and the
left MCAs, and CO$_2$ (see Table~\ref{table1}); (ii) gain, phase and
coherence from transfer function analysis (see Table~\ref{table2});
and (iii) the direct cross-correlation, the phase synchronization,
and the cross-correlation of instantaneous phase increments of BP
and BFV (see Table~\ref{table3}). We analyze BFV on the stroke side
in the right MCA in 5 patients and in the left MCA in 8 patients. In
our comparative tests, we consider the side opposite to the stroke
side as the ``normal'' side, as we did not {\it a priori} know
whether the side opposite to the stroke side would exhibit normal or
perturbed behavior. In the group comparison, we compare the stroke
side BFV for the stroke group to the BFV in the left side MCA for
the control group because the majority of post-stroke subjects had
stroke on their left side. We note that our comparative tests
between the stroke side BFV for post-stroke subjects and the right
side BFV for the control group show similar results. We compare the
normal side (opposite to the stroke side) BFV for the stroke group
to the BFV in the right side MCA for the control group.


\medskip
\noindent {\it Method 1: Transfer Function Analysis}
\medskip

We first normalize the BP and BFV signals to unit standard deviation
to obtain the respective signals $P(t)$ and $V(t)$. We then
calculate their respective Fourier transforms $V(f)$ and $P(f)$. In
the frequency domain, the coherence function $\gamma^2(f)$ is
defined as
\begin{equation}
\gamma^2(f)\equiv\frac{|S_{PV}(f)|^2}{S_{PP}(f)S_{VV}(f)},
\label{eqn1}
\end{equation}
where $S_{VV}(f)=|V(f)|^2$, $S_{PP}(f)=|P(f)|^2$ and
$S_{PV}(f)=P^*(f)V(f)$ are the power spectra of $V(t)$, $P(t)$ and
the cross-spectrum of $V(t)$ and $P(t)$, respectively. The value of
the coherence function $\gamma^2(f)$ varies between 0 and 1. The
transfer function $H(f)$ is defined as
\begin{equation}
H(f)\equiv\frac{S_{PV}(f)}{S_{PP}(f)}. \label{eqn2}
\end{equation}

From the real part $H_R(f)$  and imaginary part $H_I(f)$ of the
transfer function, we can obtain its amplitude (also called gain)
\begin{equation}
|H(f)|\equiv[H_R^2(f)+H_I^2(f)]^{1/2}, \label{eqn2a}
\end{equation}
and its phase
\begin{equation}
\Phi(f)\equiv{\rm arctan}[H_I(f)/H_R(f)]. \label{eqn2b}
\end{equation}
 We note that the phase
$\Phi(f)$ is a frequency domain characteristic of the cross-spectrum
between two signals, and is different from the instantaneous phase
in the time domain we discuss in {\it  Method 3} and {\it  Method 4}.

\begin{figure*}
\centerline{
\epsfysize=0.9\columnwidth{\rotatebox{-90}{\epsfbox{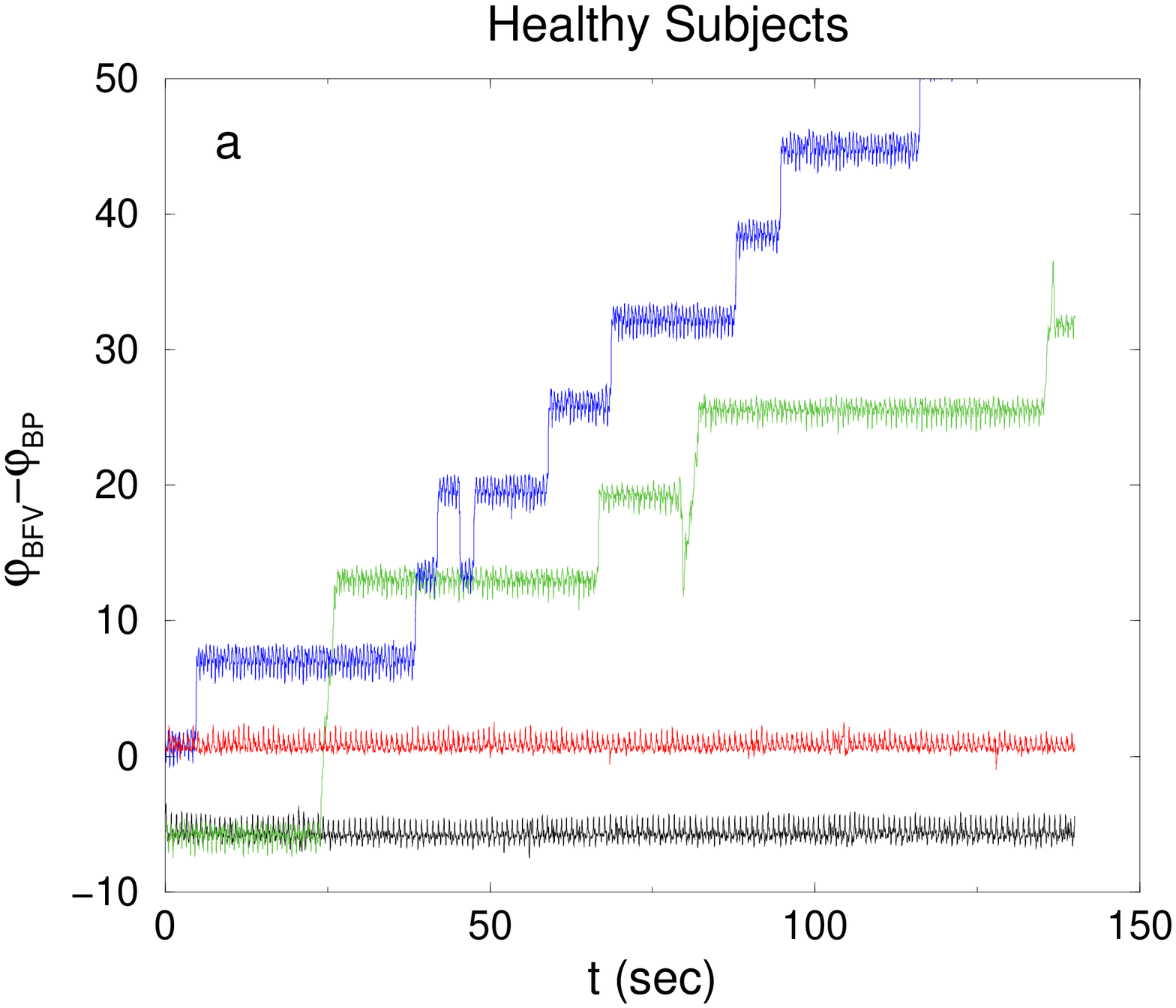}}}
\hspace{0.3cm}
\epsfysize=0.9\columnwidth{\rotatebox{-90}{\epsfbox{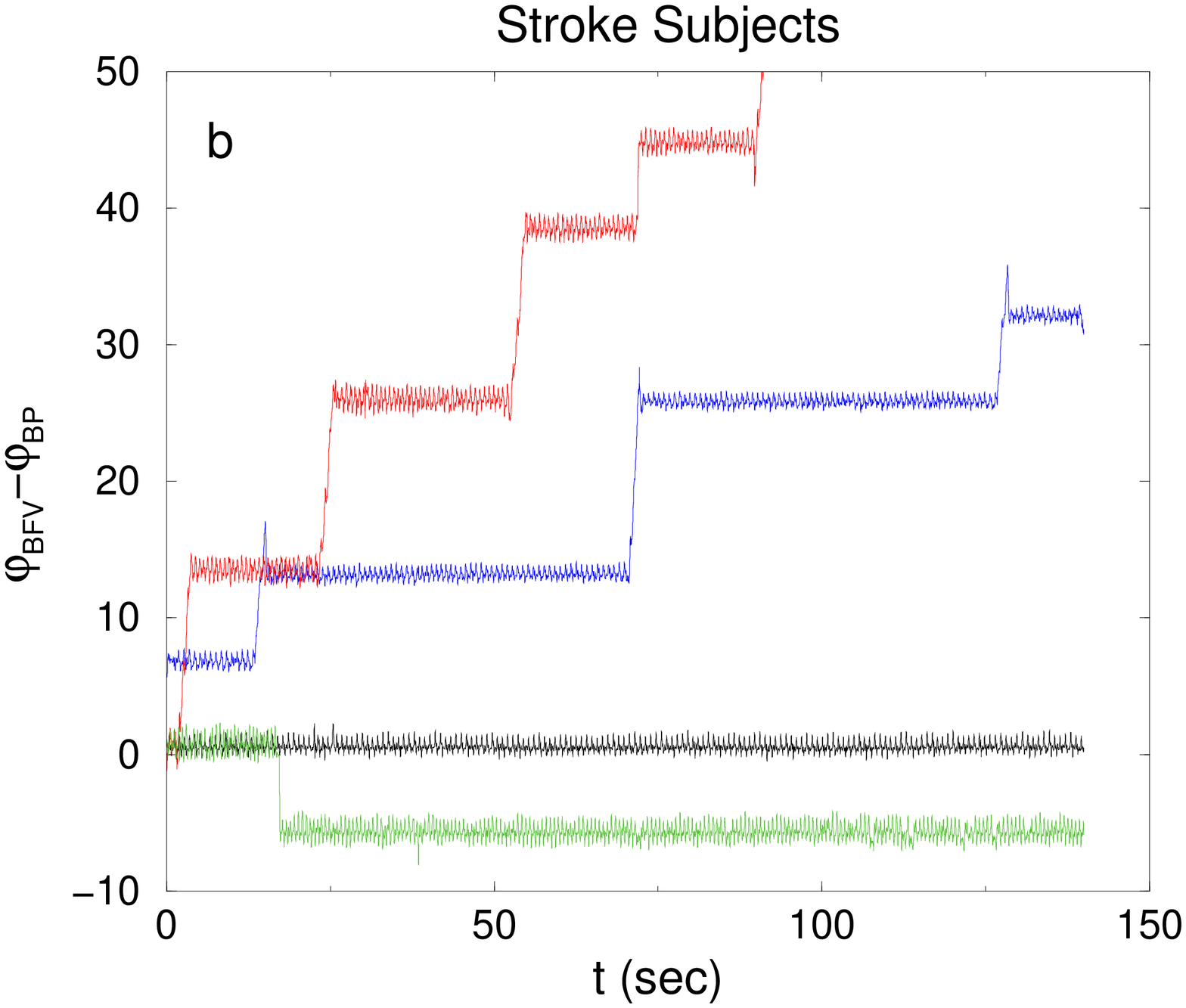}}}}
\centerline{
\epsfysize=0.9\columnwidth{\rotatebox{-90}{\epsfbox{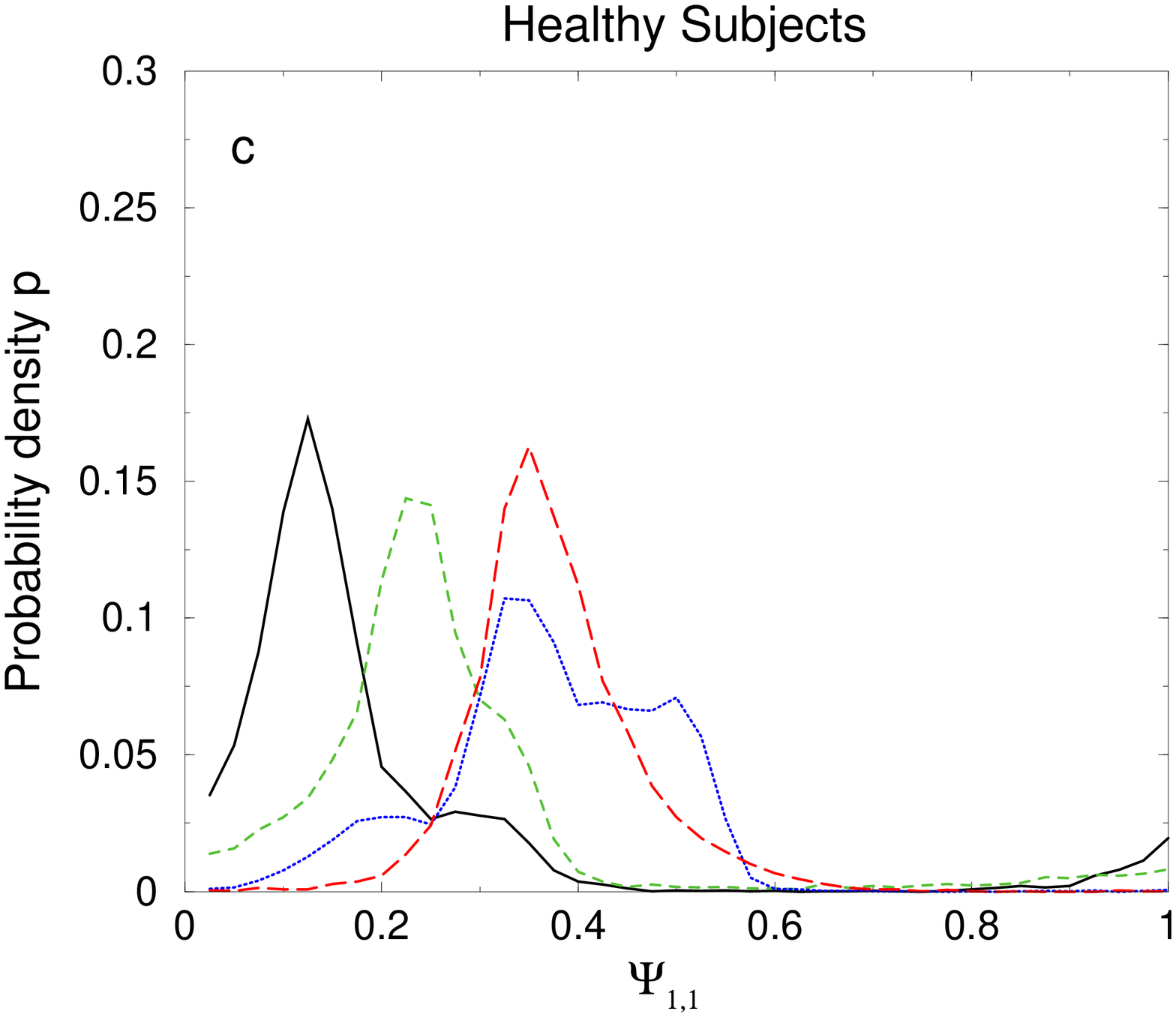}}}
\hspace{0.3cm}
\epsfysize=0.9\columnwidth{\rotatebox{-90}{\epsfbox{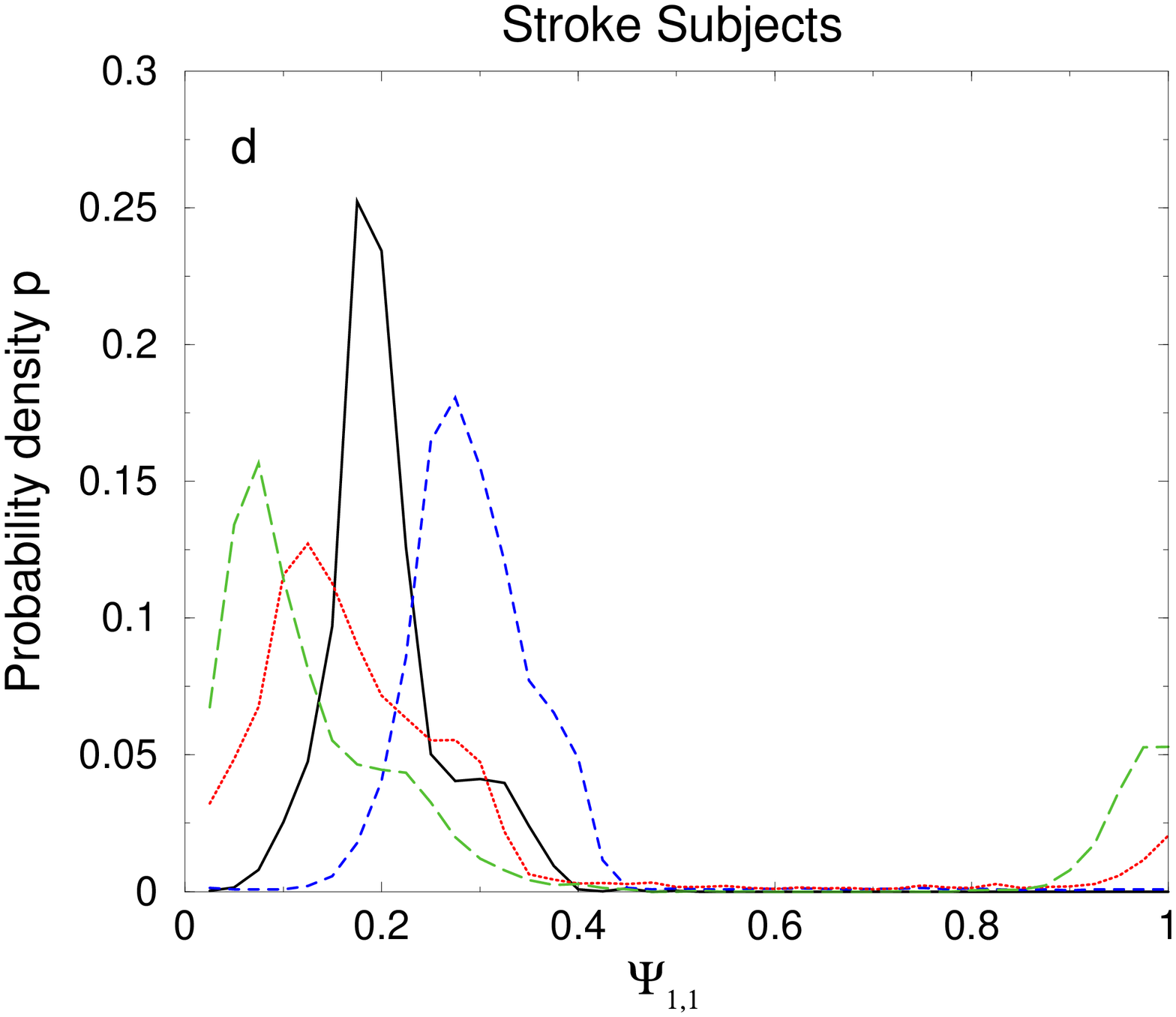}}}}
\caption{(a) The relative phases $\varphi_{BFV}(t)-\varphi_{BP}(t)$
for healthy subjects (a) and post-stroke subjects (b), where
$\varphi(t)$ is the instantaneous phase of a signal. The phase
difference of BFV and BP for both groups may fluctuate around a
constant value or jump between different constant values. The
distributions of
$\Psi_{1,1}\equiv(2\pi)^{-1}[\varphi_{BFV}(t)-\varphi_{BP}(t)]$ mod 1
for healthy subjects and for post-stroke subjects are shown in (c) and
(d), respectively. The number of bins in the histogram is $N=40$.}
\label{fig01n2}
\end{figure*}

\begin{figure*}
\epsfysize=0.9\columnwidth{\rotatebox{-90}{\epsfbox{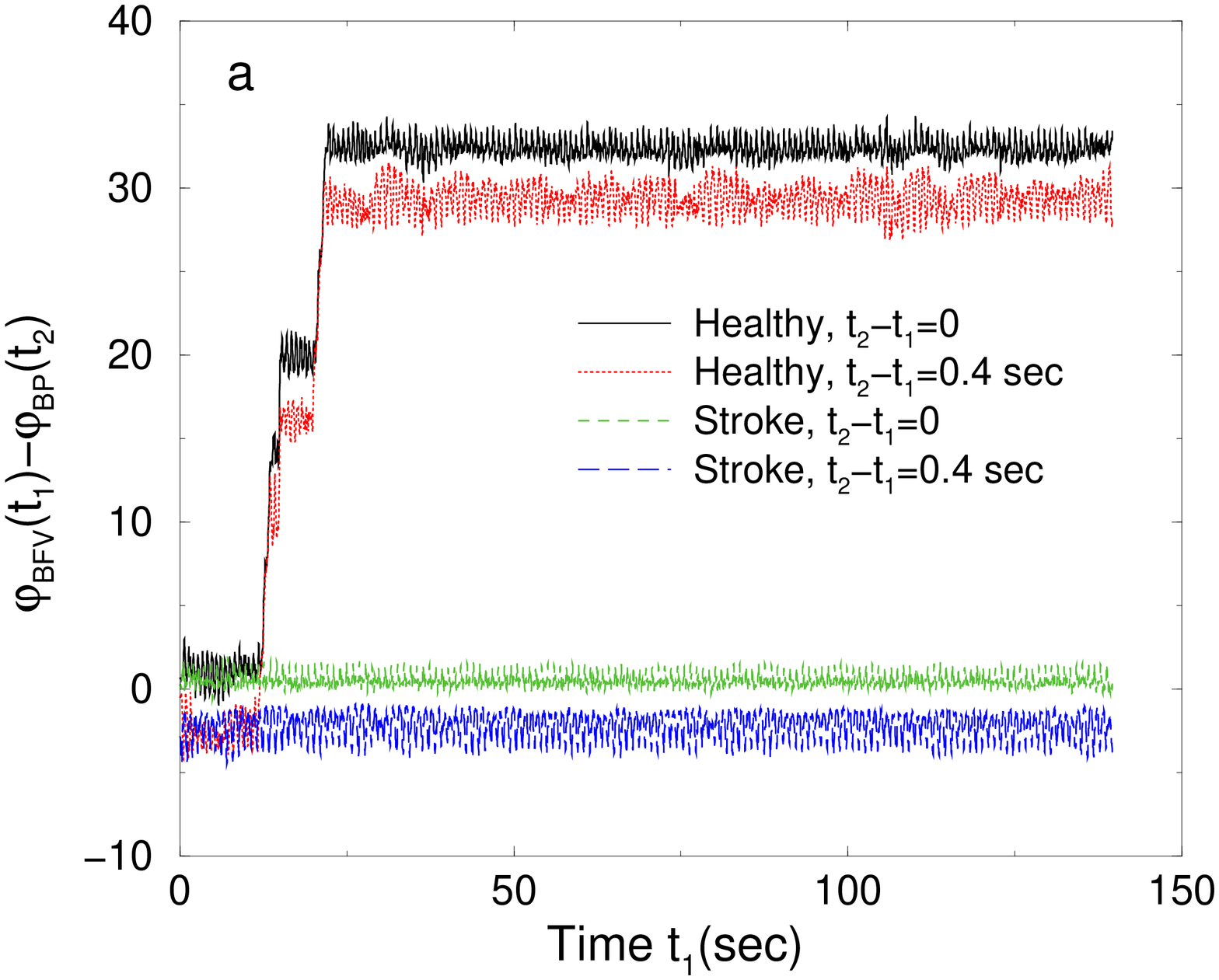}}}
\centerline{
\epsfysize=0.9\columnwidth{\rotatebox{-90}{\epsfbox{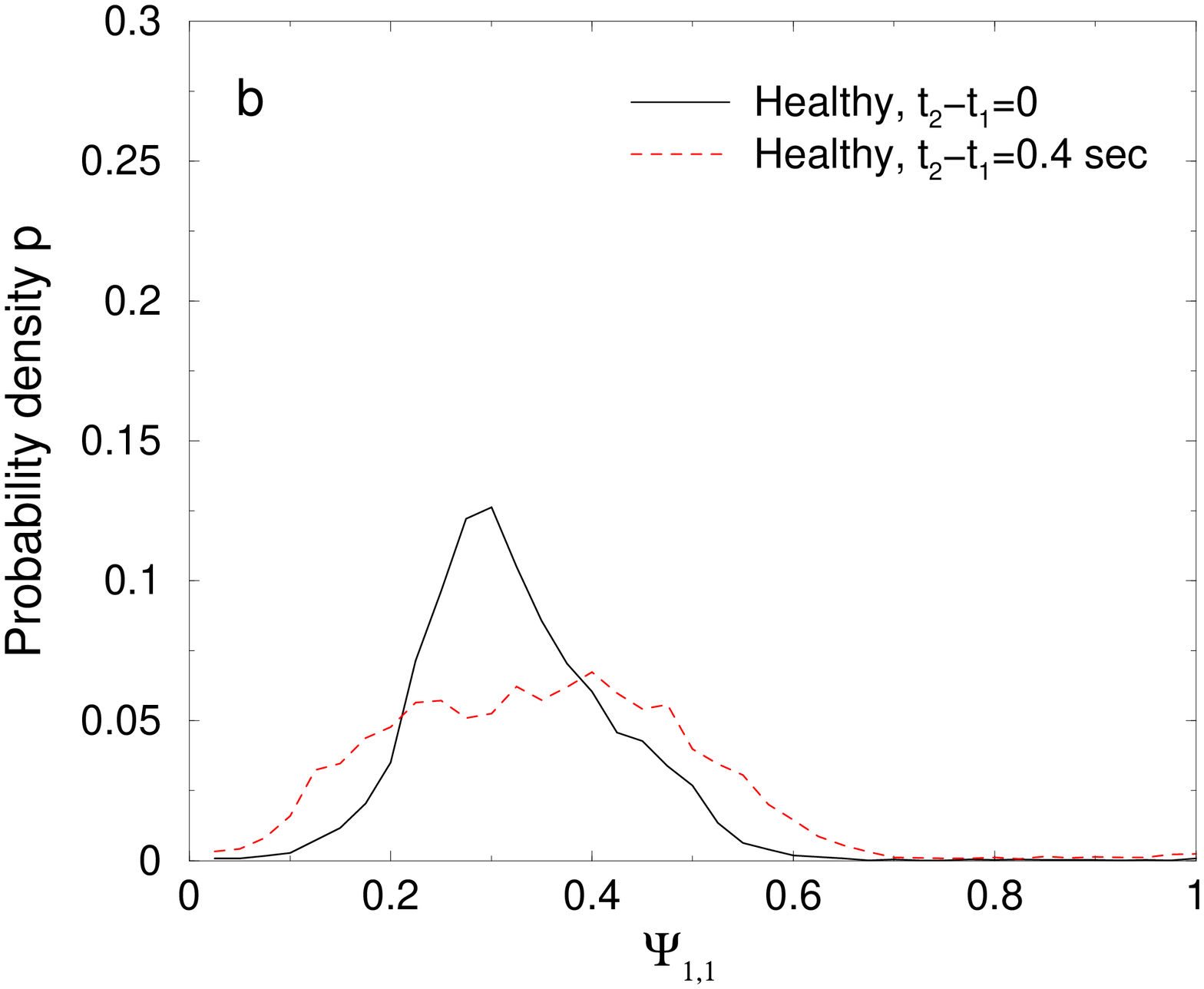}}}
\hspace{0.3cm}
\epsfysize=0.9\columnwidth{\rotatebox{-90}{\epsfbox{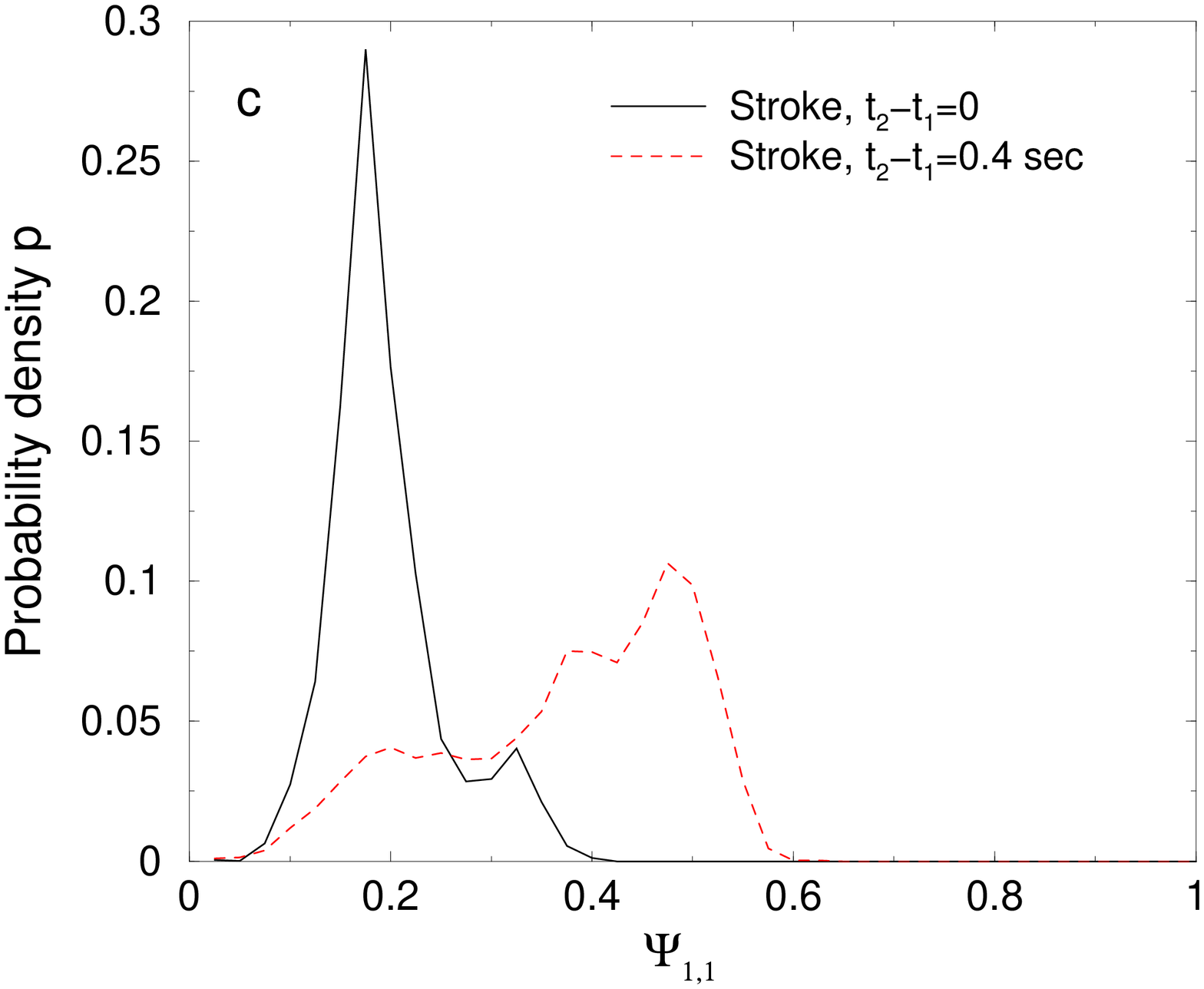}}}}
\caption{(a) The relative phases
$\varphi_{BFV}(t_1)-\varphi_{BP}(t_2)$ for a healthy subject and a
post-stroke subject. We find that the relative phases between BFV
and BP depend on the time difference $t_2-t_1$ between these two
signals. Accordingly, the distributions of
$\Psi_{1,1}\equiv(2\pi)^{-1}[\varphi_{BFV}(t_1)-\varphi_{BP}(t_2)]$
mod 1, shown in (b) and (c) for a healthy subject and a post-stroke
subject respectively, also depend on the time difference between BFV
and BP. We note that, our choice of time difference $t_2-t_1=0.4$
sec is arbitrary and does not carry any specific physiologic
meaning. The number of bins in the histogram is $N=40$.}
\label{fig01n3}
\end{figure*}

\medskip
{\it \noindent Method 2: Cross-correlation Analysis}
\medskip

To test dynamical aspects of the mechanism of CA, we investigate the
cross-correlation between BP and BFV signals and how this
cross-correlation changes under stroke. Although we consider
segments of the BP and BFV recordings during the quasi-steady state,
where we have constant level (or absence) of physiologic stimuli,
these signals may still exhibit certain trends in the mean value.
So, to eliminate these trends, we first pre-process the BP and BFV
signals applying a band-pass filter ($10{\rm Hz}>f>0.05$Hz) in the
frequency domain. To be able to compare signals with different
amplitude, we next normalize the band-passed signals to unit
standard deviation (Fig.~\ref{fig01}). Finally, we perform a
cross-correlation analysis estimating the cross-correlation function
$C(\tau)$ for a broad range of values for the time lag $\tau$, where
$C(\tau)$ is defined as:
\begin{equation}
C(\tau)\equiv\frac{\langle(P(t)-\langle P\rangle)(V(t+\tau)-\langle
V\rangle)\rangle}{\sigma_P\sigma_V}, \label{eqn2n}
\end{equation}
and $\langle...\rangle$ and $\sigma$ denote the mean and standard
deviation of a signal.

Results of our cross-correlation analysis for one healthy subject
and one post-stroke subject during all four physiologic conditions
are shown in Fig.~\ref{fig01n} and are discussed in
Sec.~\ref{res_corr}.


\medskip
\noindent {\it Method 3: Phase Synchronization Analysis}
\medskip

We study beat-to-beat BP-BFV interaction during quasi-steady state
conditions (supine rest, upright tilt, upright hyperventilation and
CO$_2$ rebreathing employing a phase synchronization method. We first
apply a high-pass ($f>0.05$Hz) and a low-pass ($f<10$Hz) Fourier
filter to the BP and BFV signals. The high-pass filter is used to
reduce nonstationarity related to slow trends in the mean of the
signals. The low-pass filter is used to remove high frequency random
fluctuations in the signals. Next we perform the Hilbert transform
which for a time series $s(t)$ is defined as
~\cite{Rosenblum96,Rosenblum97,Claerbout76,Oppenheim98,Plamenbook98,Marple99}
\begin{equation}
\tilde
s(t)\equiv\frac{1}{\pi}P\int^{\infty}_{-\infty}\frac{s(\tau)}{t-\tau}d\tau,
\label{eqn3}
\end{equation}
where P denotes Cauchy principal value.  $\tilde s(t)$ has an
apparent physical meaning in Fourier space: for any positive
(negative) frequency $f$, the Fourier component of the Hilbert
transform $\tilde s(t)$  at this frequency $f$ can be obtained from
the Fourier component of the original signal $s(t)$ at the same
frequency $f$ after a 90$^{\circ}$ clockwise (anti-clockwise)
rotation in the complex plane. For example, if the original signal
is ${\rm sin}(\omega t)$, its Hilbert transform will become ${\rm
cos}(\omega t)$. For any signal $s(t)$ one can always construct its
``analytic signal''
$S$~\cite{Rosenblum96,Rosenblum97,Claerbout76,Oppenheim98,Marple99},
which is defined as
\begin{equation}
S\equiv s(t)+i\tilde s(t)=A(t)e^{i\varphi(t)}, \label{eqn4}
\end{equation}
where $A(t)$ and $\varphi(t)$ are the instantaneous amplitude and
instantaneous phase of $s(t)$, respectively. Application of the
analytic signal approach to heartbeat dynamics has been shown
in~\cite{Plamennature96,Plamenbook98}. The instantaneous amplitude
$A(t)$ and the instantaneous phase $\varphi(t)$ are instantaneous
characteristics of a time series $s(t)$, and present different
aspects of the signal. For a pure sinusoid, $A(t)$ is a constant and
$\varphi(t)$ increases linearly in time: the amplitude quantifies
the strength of the oscillation and the slope of the straight line
formed by the increasing phase quantifies how fast is the
oscillation. For more complex signals, both $A(t)$ and $\varphi(t)$
may display complicated forms. Further, we note that the
instantaneous phase $\varphi(t)$ is different from the transfer
function phase $\Phi(f)$: $\varphi(t)$ is a time domain
characteristic of a single signal, while $\Phi(f)$ is a
cross-spectrum characteristic of two signals in the frequency
domain.

From the definition of the Fourier transform one can find that the
mean of a given signal $s(t)$ is proportional to the Fourier
component of $s(t)$ at the frequency $f=0$. After applying the
high-pass filter ($f>0.05$ Hz), the Fourier component of $s(t)$ at
$f=0$ is filtered out, and correspondingly, the mean of the filtered
signal becomes zero. Furthermore, from the definition of the Hilbert
transform, one can find that the Hilbert transforms of two signals
$s(t)$ and $s(t)+c$ (where $c\ne0$ is a constant) are identical,
although the two original signals have obviously different means due
to the constant $c$. Since both the instantaneous phase $\varphi(t)$
and amplitude $A(t)$ depend on the original signal as well as on the
Hilbert transform of the original signal (see Eq.~\ref{eqn4}), the
instantaneous phase and amplitude for $s(t)$ and $s(t)+c$ will be
also different. However, when we let both signals $s(t)$ and
$s(t)+c$ pass a high-pass filter ($f>0.05$ Hz), the mean of both
signals becomes zero, and both signals will have identical phases
and amplitudes. Thus, the phase and amplitude of a signal does not
depend on its mean, and is uniquely defined after the signal is
processed with a high-pass filter.

\begin{figure*}
\centerline{
\epsfysize=0.9\columnwidth{\rotatebox{-90}{\epsfbox{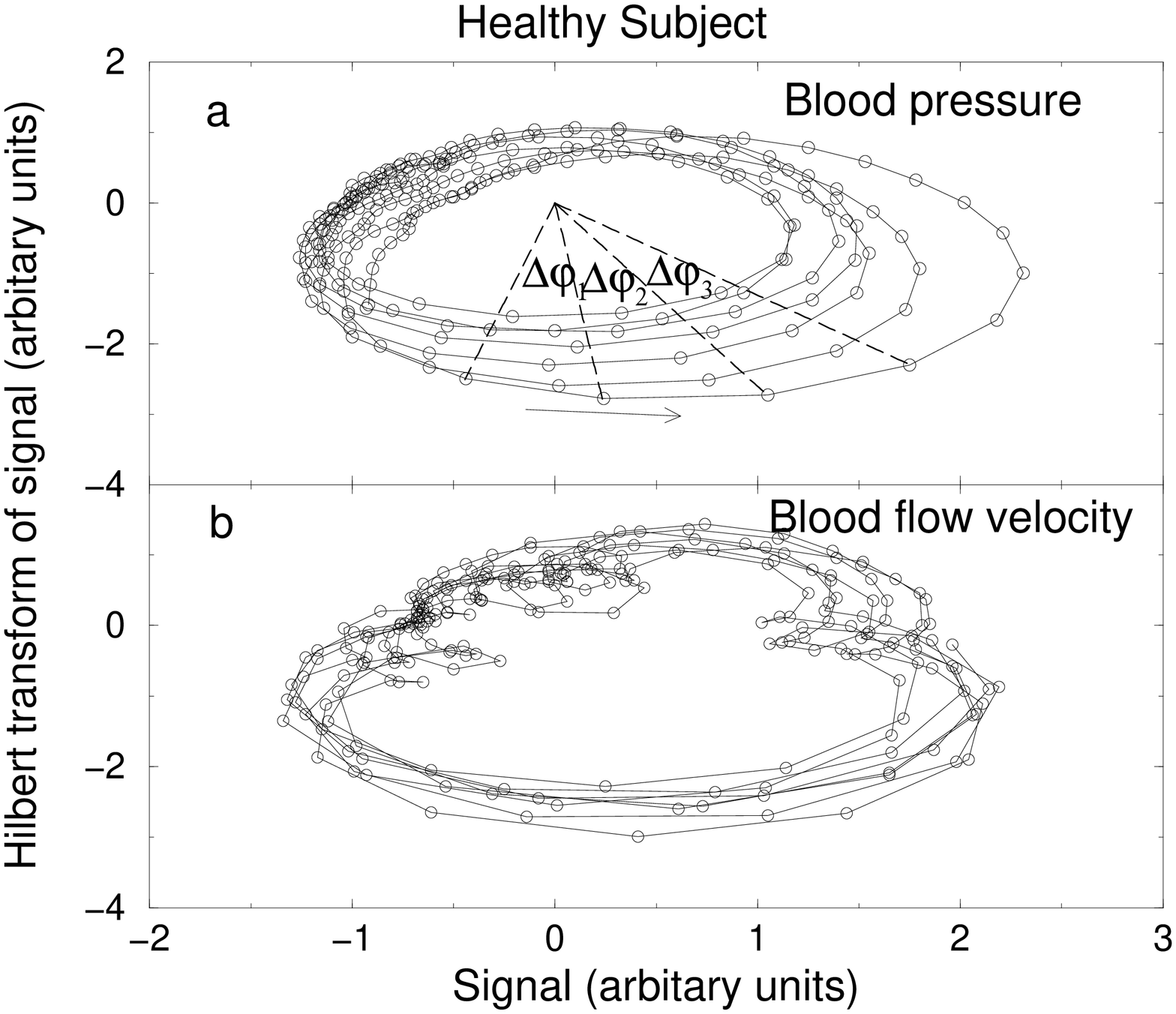}}}
\hspace{0.3cm}
\epsfysize=0.9\columnwidth{\rotatebox{-90}{\epsfbox{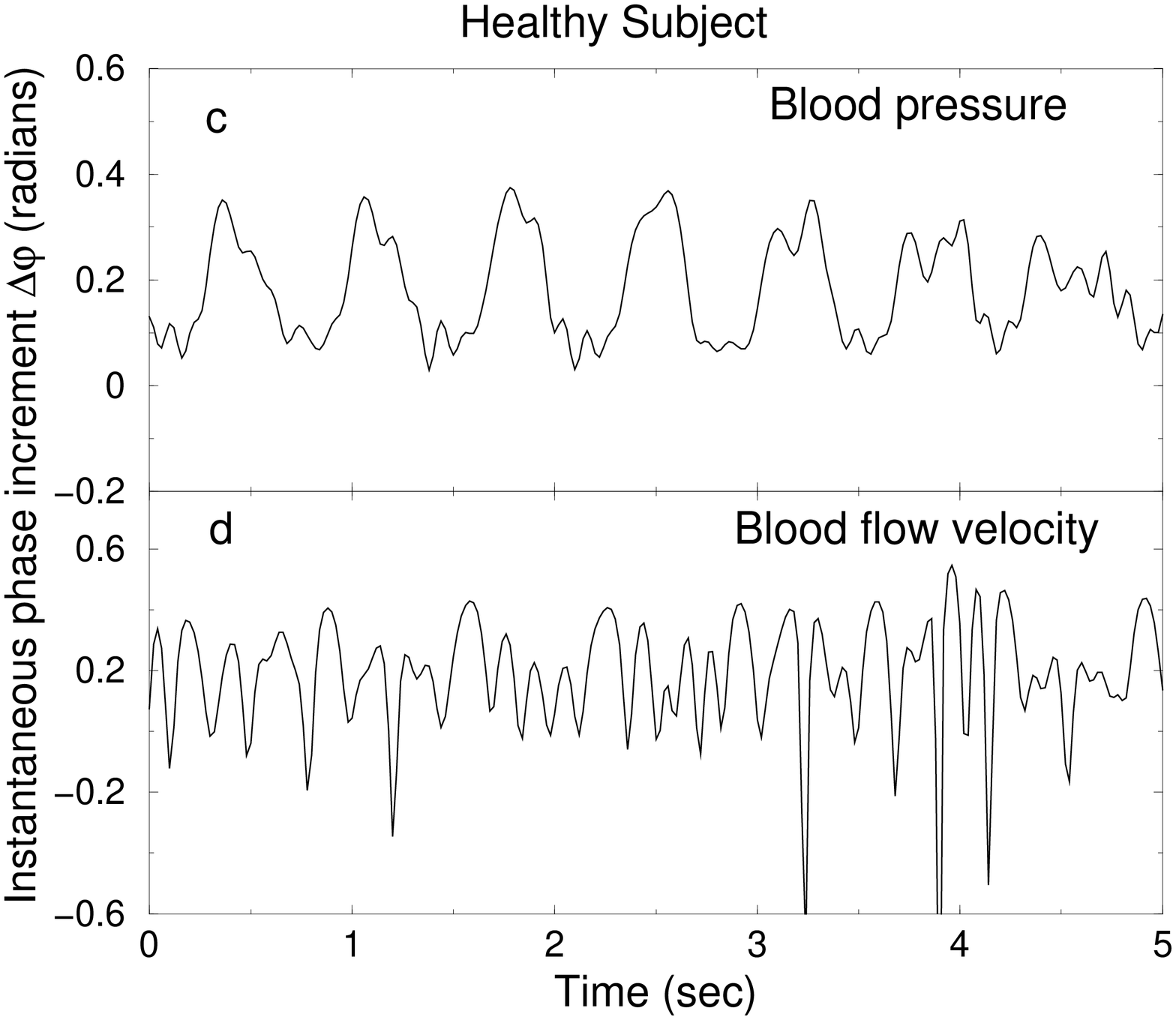}}}}
\centerline{
\epsfysize=0.9\columnwidth{\rotatebox{-90}{\epsfbox{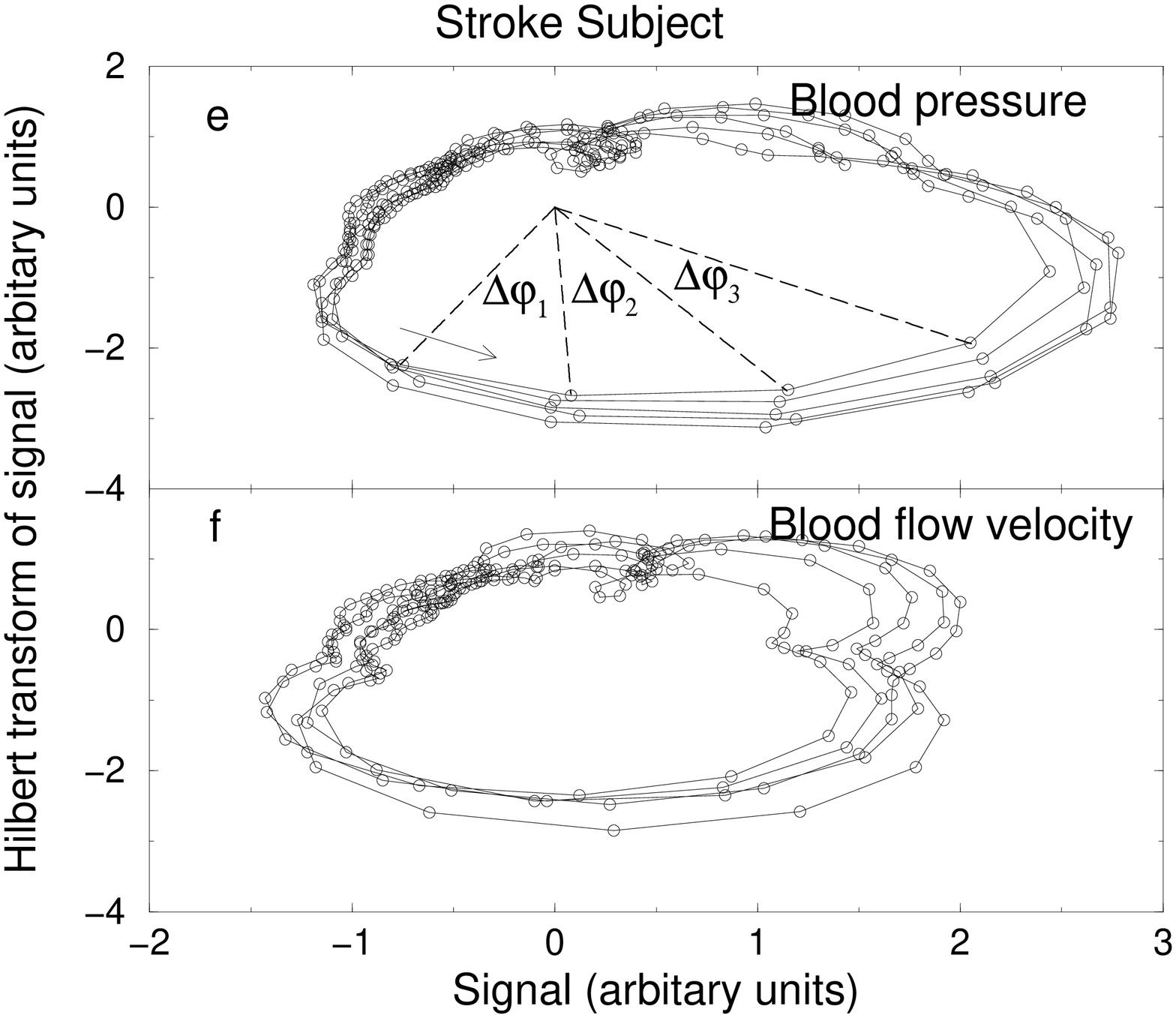}}}
\hspace{0.3cm}
\epsfysize=0.9\columnwidth{\rotatebox{-90}{\epsfbox{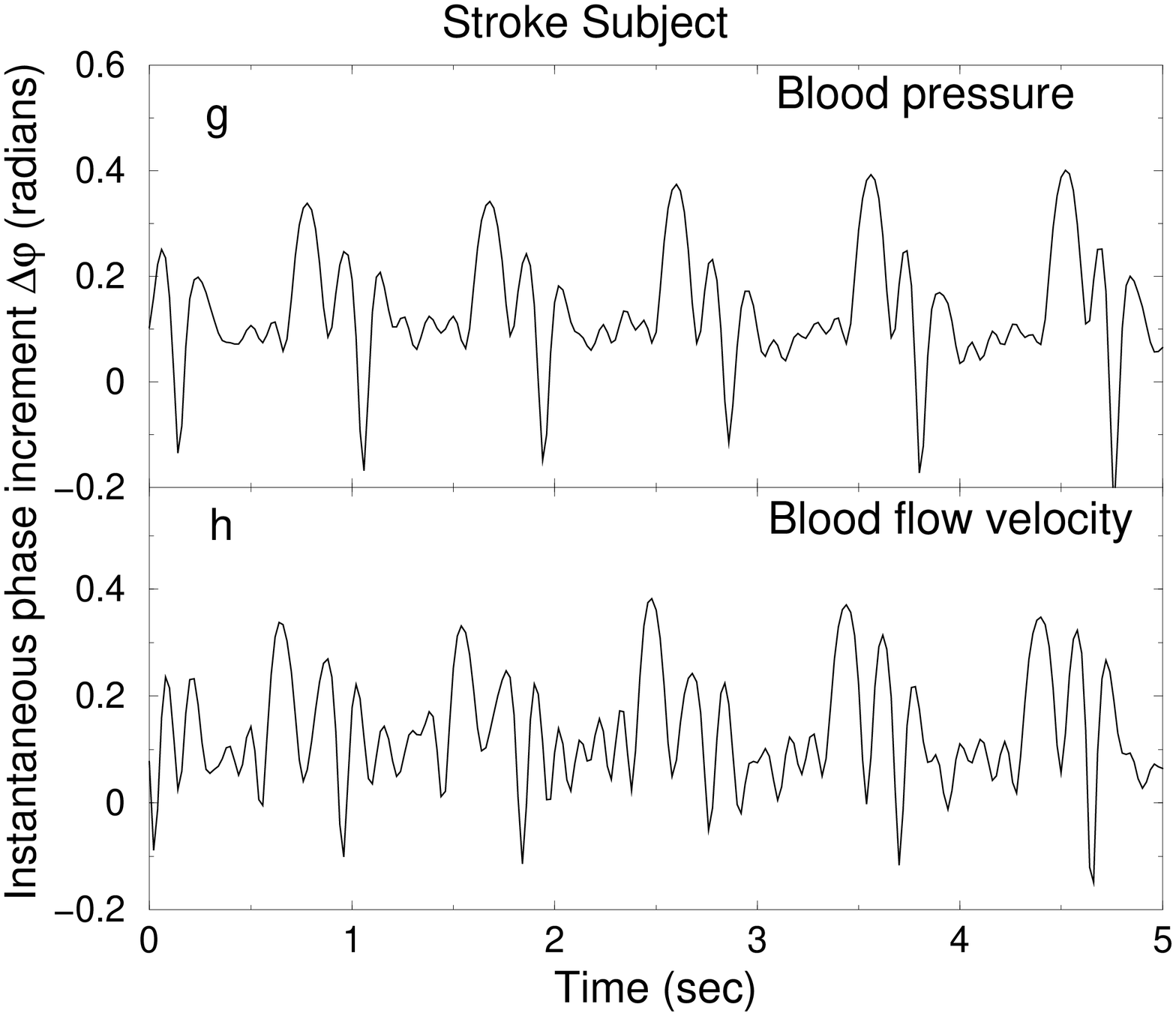}}}}
\caption{Presentation of the BP and BFV signals vs. their Hilbert
transforms (a-b) and their corresponding instantaneous phase
increment $\Delta \varphi$ during the CO$_2$ re-breathing condition
(c-d) for the same data from a healthy subject as shown in
Fig.~\ref{fig01}a-b. BP and BFV signals vs. their Hilbert transforms
(e-f) and their corresponding instantaneous phase increment $\Delta
\varphi$ during the CO$_2$ re-breathing condition (g-h) for the same
data from a post-stroke subject as shown in Fig.~\ref{fig01}c-d.
Repetitive temporal patterns associated with each heartbeat in
$\Delta \varphi$ for the BP signal from a healthy subject (c) are
not matched by corresponding patterns in the BFV signal (d),
reflecting active cerebral vascular regulation. In contrast,
periodic patterns in $\Delta \varphi$ of the BP signal from a
post-stroke subject (g) are matched by practically identical
patterns in $\Delta \varphi$ of the BFV signal (h), indicating
dramatic impairment of cerebral vascular tone with higher vascular
resistance after minor ischemic stroke.} \label{fig02}
\end{figure*}

Following \cite{Rosenblum96} we estimate the difference between the
instantaneous phases of the BFV and BP signals
$\varphi_{BFV}(t)-\varphi_{BP}(t)$ at the same time $t$. We then
obtain the probability density $p$ for the quantity
\begin{equation}
\Psi_{1,1}\equiv(2\pi)^{-1}[\varphi_{BFV}(t)-\varphi_{BP}(t)]\mbox{~mod~1},
\label{eqn4n0}
\end{equation}
following the approach presented in \cite{Tass98}. To quantify the
shape of the probability density $p$, which can be very broad
(uniform) for non-synchronized signals and very narrow (peaked) for
strongly synchronized signals, we utilize the index
$\rho$~\cite{Tass98}. The index $\rho$ is based on the Shannon
entropy and is defined as
\begin{equation}
\rho\equiv(S_{\rm max}-S)/S_{\rm max}, \label{eqn4n}
\end{equation}
where $S=-\sum^N_{k=1}p_k {\rm ln } p_k$ is the Shannon entropy of
the distribution of $\Psi_{1,1}$ and $S_{\rm max}={\rm ln }N$
corresponds to the uniform distribution, where $N$ is the number of
bins. In our calculations, we choose $N=40$. For the normalized
index $\rho$, we have $0\leq \rho \leq 1$, where $\rho=0$
corresponds to a uniform distribution (no synchronization) and
$\rho=1$ corresponds to a Dirac-like distribution (perfect
synchronization).

Results of our synchronization analysis for several healthy and
post-stroke subjects are shown in Figs.~\ref{fig01n2} and
\ref{fig01n3}, and are discussed in Sec.~\ref{res_syn}.

\medskip
\noindent {\it Method 4: Cross-correlation of Instantaneous Phase
Increments}
\medskip

The instantaneous phase $\varphi(t)$ for both BP and BFV is a
nonstationary signal and can be decomposed into two parts: a linear
trend and fluctuations along the trend. The trend is mainly driven
by the periodic heart rate at a frequency $\approx$ 1 Hz. However,
the fluctuations are of specific interest, since they may be
affected by the cerebral autoregulation. To remove the trend, we
consider the increments in the consecutive values of the
instantaneous phase, defined as
\begin{equation}
\Delta \varphi(t_i)\equiv\varphi(t_i)-\varphi(t_{i-1}),
\label{eqn4n2}
\end{equation}
where $t_i$ and $t_{i-1}$ are the times corresponding to two
successive recordings (in our case we have $t_i-t_{i-1}=0.02$sec).
The instantaneous phase increment signal $\Delta \varphi$ is
stationary in the sense that it has fixed mean and fixed standard
deviation, and fluctuates in the range $(-\pi,\pi]$. In
Figs.~\ref{fig02}(c),~\ref{fig02}(d) we show examples of $\Delta
\varphi$ for healthy subjects and in
Figs.~\ref{fig02}(g),~\ref{fig02}(h) for post-stroke subjects.

\begin{table*}
{ \scriptsize
\begin{tabular}{c|c|c|cc|cc|cc|cc|cc}
\hline Frequency & Variable & &\multicolumn{2}{c|}{Supine}  &
\multicolumn{2}{c|}{Tilt} & \multicolumn{2}{c|}{Hyperventilation}
&\multicolumn{2}{c|}{CO$_2$ rebreathing} &
\multicolumn{2}{c}{Statistics}\\\cline{4-13}

Range & & & BFV- & BFV- & BFV-&BFV- & BFV- & BFV-& BFV- & BFV-&
BFV- & BFV-\\

& & & MCAR/ & MCAL/ &MCAR/ &MCAL/ &MCAR/ &MCAL/ &MCAR/ &MCAL/ &MCAR/
&MCAL/\\

& & & Normal & Stroke &Normal &Stroke &Normal &Stroke &Normal
&Stroke &Normal &Stroke\\

& & & side & side & side & side & side & side & side & side & side &
side\\\hline

Low& Gain & Control & 0.94$\pm$ 0.30& 0.97$\pm$0.27& 1.06$\pm$0.21 &
$0.97\pm0.21$& $0.95\pm0.41$& $1.02\pm0.43$& $1.03\pm0.15$&
$1.03\pm0.15$
& 0.38$^{*}$& 0.11$^{*}$\\
Frequency& Gain & Stroke  & 0.84$\pm$ 0.42& 0.72$\pm$0.37&
0.95$\pm$0.27& $0.95\pm0.25$& $1.09\pm0.36$& $1.08\pm0.38$&
$0.92\pm0.20$& $0.95\pm0.21$& 0.48$^{\dagger}$&0.26$^{\dagger}$
\\\cline{2-13}

0.05-0.2Hz& Coherence & Control & 0.65$\pm$ 0.15& 0.59$\pm$0.21&
0.75$\pm$0.08 & $0.66\pm0.19$& $0.58\pm0.18$& $0.62\pm0.19$&
$0.82\pm0.09$& $0.84\pm0.08$
& 0.0006$^{*}$& 0.0004$^{*}$\\
& Coherence & Stroke  & 0.53$\pm$ 0.19& 0.44$\pm$0.24&
0.70$\pm$0.20& $0.66\pm0.27$& $0.62\pm0.17$& $0.61\pm0.19$&
$0.69\pm0.19$& $0.70\pm0.20$& 0.07$^{\dagger}$&0.07$^{\dagger}$
\\\cline{2-13}

& Phase (rad) & Control & 0.56$\pm$ 0.27& 0.63$\pm$0.26&
0.62$\pm$0.32 & $0.63\pm0.30$& $0.92\pm0.33$& $0.94\pm0.39$&
$0.63\pm0.23$& $0.63\pm0.20$
& 0.0002$^{*}$& 0.0001$^{*}$\\
& Phase (rad) & Stroke  & 0.79$\pm$ 0.27& 0.56$\pm$0.38&
0.58$\pm$0.24& $0.60\pm0.25$& $0.86\pm0.32$& $0.94\pm0.47$&
$0.42\pm0.25$& $0.49\pm0.23$& 0.74$^{\dagger}$&0.37$^{\dagger}$
\\\hline

Heartbeat& Gain & Control & 0.92$\pm$ 0.18& 0.93$\pm$0.24&
0.91$\pm$0.15 & $0.91\pm0.21$& $0.88\pm0.23$& $0.94\pm0.23$&
$0.89\pm0.16$& $0.91\pm0.18$
& 0.50$^{*}$& 0.73$^{*}$\\
Frequency& Gain & Stroke  & 0.77$\pm$ 0.29& 0.82$\pm$0.28&
0.88$\pm$0.27& $0.97\pm0.43$& $0.97\pm0.24$& $0.98\pm0.28$&
$0.96\pm0.28$& $0.97\pm0.31$& 0.96$^{\dagger}$&0.84$^{\dagger}$
\\\cline{2-13}

0.7-1.4Hz& Coherence & Control & 0.76$\pm$ 0.15& 0.75$\pm$0.18&
0.75$\pm$0.13 & $0.71\pm0.18$& $0.58\pm0.16$& $0.58\pm0.18$&
$0.68\pm0.23$& $0.70\pm0.22$
& 0.15$^{*}$& 0.27$^{*}$\\
& Coherence & Stroke  & 0.64$\pm$ 0.24& 0.63$\pm$0.25&
0.57$\pm$0.23& $0.56\pm0.21$& $0.55\pm0.20$& $0.56\pm0.21$&
$0.64\pm0.25$& $0.63\pm0.27$& 0.035$^{\dagger}$&0.05$^{\dagger}$
\\\cline{2-13}

& Phase (rad) & Control & 0.32$\pm$ 0.13& 0.27$\pm$0.14&
0.38$\pm$0.18 & $0.38\pm0.20$& $0.46\pm0.27$& $0.46\pm0.26$&
$0.46\pm0.26$& $0.42\pm0.24$
& 0.16$^{*}$& 0.03$^{*}$\\
& Phase (rad) & Stroke  & 0.35$\pm$ 0.31& 0.31$\pm$0.24&
0.36$\pm$0.14& $0.36\pm0.21$& $0.47\pm0.20$& $0.49\pm0.28$&
$0.42\pm0.21$& $0.42\pm0.15$& 0.95$^{\dagger}$&0.81$^{\dagger}$
\\\hline
\end{tabular}
}

* P value between physiologic conditions comparisons

$\dagger$ P value between groups comparisons

\caption{Gain, coherence and phase in the low frequency (LF,
0.05-0.2 Hz) range and in the heartbeat frequency (HBF, 0.7-1.4 Hz)
range for the control and stroke group during different physiologic
conditions. We compare data from BFV in the right MCA (BFV-MCAR) in
healthy subjects with data from BFV in the normal side MCA in
post-stroke subjects (mean $\pm$ standard deviation values are
presented in the left column for each condition). We compare data
from BFV in the left MCA (BFV-MCAL) in healthy subjects with data
from BFV in the stroke side MCA in post-stroke subjects (mean $\pm$
standard deviation values are presented in the right column for each
condition). The p values from 2x4 MANOVA are calculated for
comparing differences between groups and conditions.} \label{table2}
\end{table*}

We then apply a cross-correlation analysis to quantify the dynamical
relationship between the stationary phase increments $\Delta
\varphi$ of the BP and BFV signals. For each subject, during each
physiologic condition we calculate the correlation coefficient
$C(\tau)$ vs. the time lag $\tau$ between the BP and BFV signals. To
quantitatively distinguish the control group and the stroke group,
we further analyze the characteristics of the correlation function
$C(\tau)$. Specifically, we investigate the maximum value of
$C(\tau)$, denoted as $C_{\rm max}$, which represents the strength
of the cross-correlation between the instantaneous phases of the BP
and BFV signals. Another important characteristic of the
cross-correlation function is how fast the correlation between two
signals decreases for increasing values of the time lag $\tau$. To
quantify this aspect, we choose a fixed threshold value, $r=0.3$,
which is the same for all subjects and for which we obtain a good
separation between the control and the stroke group. Since $C(\tau)$
is a periodic-like function of the time lag $\tau$ with a decreasing
amplitude for increasing $\tau$ (Fig.~\ref{fig03}), we first record
all maxima of $|C(\tau)|$ during each heart beat period ($\sim$1
sec), then we determine the two maxima with largest positive and
negative time lags $\tau$ at which the correlation function
$C(\tau)$ is above $rC_{\rm max}$. The average of the absolute
values for these two time lags is marked as a characteristic time
lag $\tau_0$.

In summary we note that, to avoid problems with the nonstationarity
when applying the cross-correlation analysis between instantaneous
phase increments, we have performed the following steps: (1) We have
considered BFV and BP at the steady state for all four physiologic
conditions, and we have analyzed only short segments ($\approx 2.3$
minutes) of data during which the heart rate remains approximately
stable; (2) We have performed cross-correlation analysis after
pre-processing BFV and BP signals by a high-pass filter, thus
removing low frequency trends; (3) The phase increment signals
$\Delta \varphi$ are stationary at least according to the definition
of weak stationarity. For the data segments we consider, $\Delta
\varphi$ has a constant mean value and fluctuates within a fixed
range $(-\pi,+\pi]$.


\section{Results}\label{secres}

\subsection{Mean Values}
\label{res_mean}


We compare the mean values of all signals for both control and
stroke groups and for all four physiologic conditions (baseline
supine rest, upright tilt, tilt-hyperventilation and CO$_2$
rebreathing) using the MANOVA method. Our results are shown in
Table~\ref{table1}. We see that the mean values of the BFV and
CO$_2$ signals are significantly different for the four different
conditions while the BP mean values are similar. For control and
stroke group comparison, we find that BFVs from the left (stroke
side) MCA were significantly different, and that the mean value of
BP for the stroke group is significantly higher than that for
control group. Furthermore, we observe that the cerebral vascular
resistance (CVR) is significantly higher for the stroke group.

\subsection{Transfer Function Analysis}
\label{res_FFT}


We apply transfer function analysis on the original BP and BFV signals
under different physiologic conditions in both the low frequency (LF)
(0.05-0.2 Hz) and the heart beat frequency (HBF) (0.7-1.4 Hz)
range. Gain, phase and coherence are calculated for each subject and
for all four physiologic conditions (Table~\ref{table2}). We use
MANOVA to compare our results for the two groups and for the four
conditions. In both frequency ranges, we do not find significant
difference in the gain. In the LF range, phase $\Phi(f)$ and coherence
$\gamma^2(f)$ are significantly different between the physiologic
conditions, but are not different between the groups. In the HBF
range, we find that the phase $\Phi(f)$ for the MCAL-BFV is
significantly different between the conditions (p=0.03). The coherence
in the HBF range shows no significant difference in the physiologic
conditions comparison, however, it is significantly higher for the
control group.

\subsection{Cross-correlation Analysis}
\label{res_corr}

We have performed a cross-correlation analysis directly between the
BP and BFV signals, after first pre-processing these signals by a
high-pass filter as shown in Fig.~\ref{fig01}. Representative
examples for the cross-correlation function over a broad range of
time lags $\tau$ for one healthy and one post-stroke subject during
the four physiologic conditions (supine, tilt, hyperventilation,
CO$_2$ rebreathing) are shown in Fig.~\ref{fig01n}. Group statistics
and comparative tests are included in Table III.

\begin{figure*}
\centerline{
\epsfysize=0.95\columnwidth{\rotatebox{-90}{\epsfbox{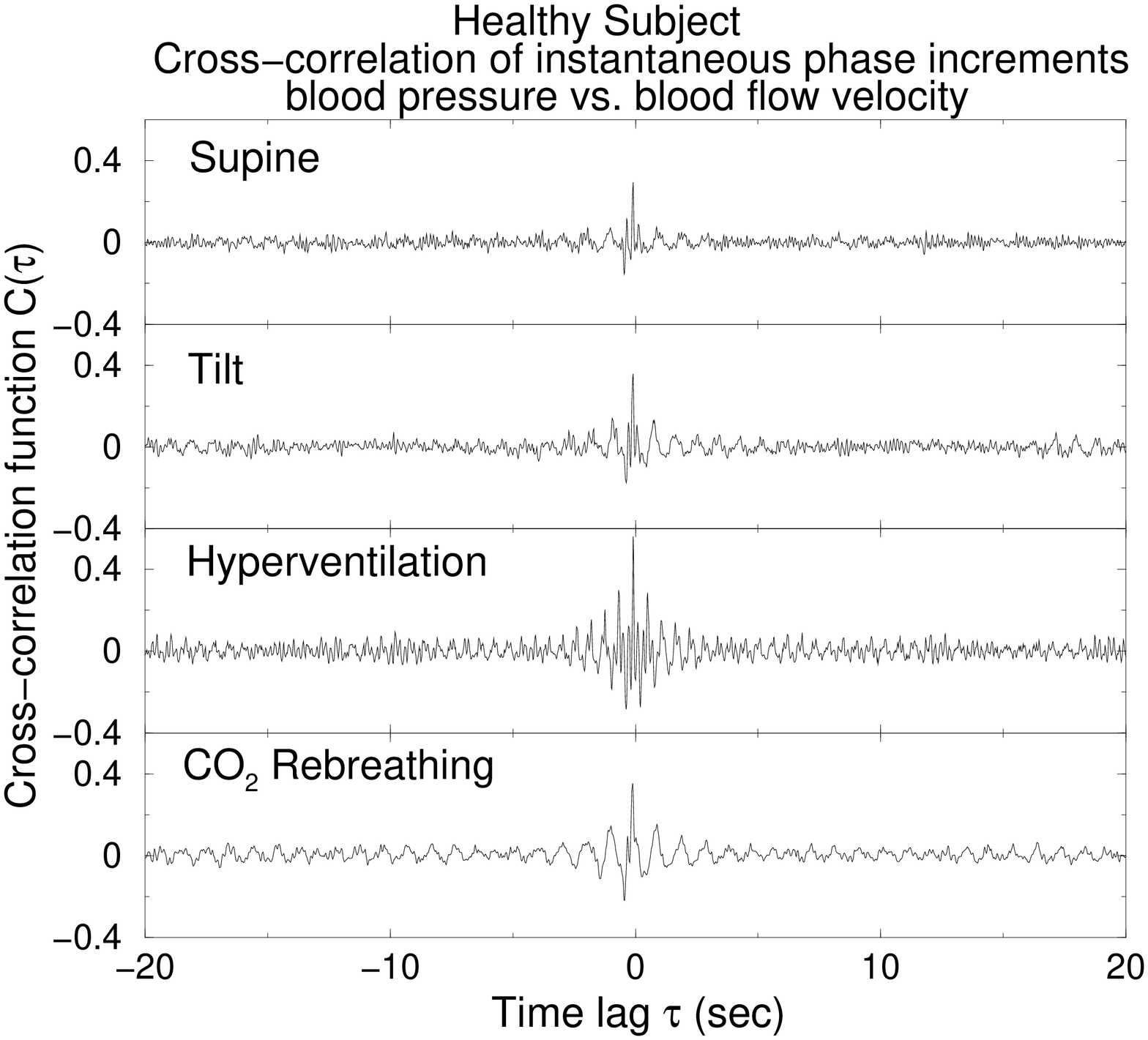}}}
\hspace{0.3cm}
\epsfysize=0.95\columnwidth{\rotatebox{-90}{\epsfbox{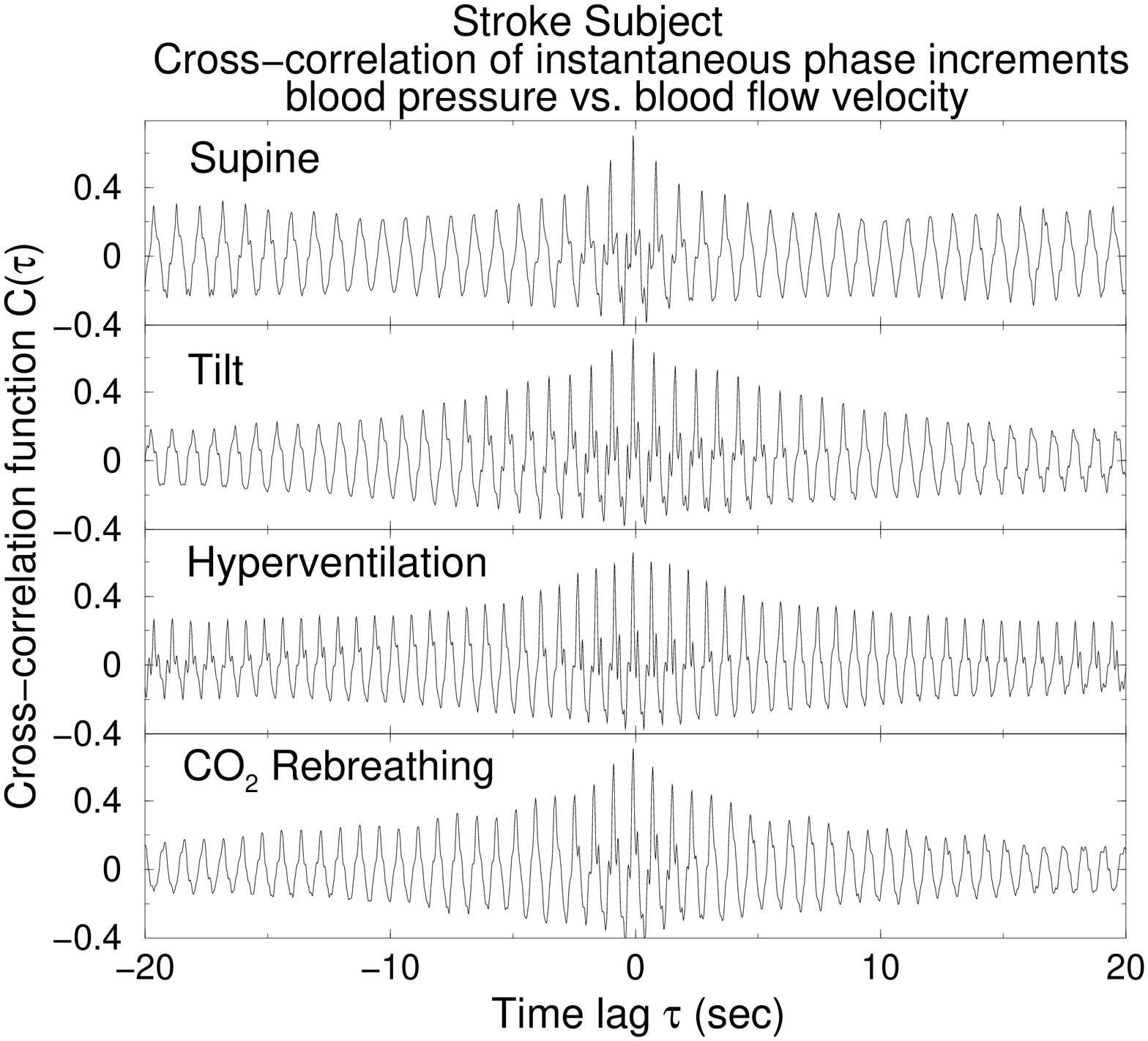}}}}
\caption{Cross-correlation function $C(\tau)$ of the instantaneous
phase increment $\Delta \varphi$ for the BP and BFV signals during
four physiologic conditions. We find that the cross-correlation
function for all healthy subjects exhibits a very distinct type of
behavior compared to post-stroke subjects. Two typical examples are
shown. (Left) A healthy subject: $C(\tau)$ has a small amplitude at
$\tau=0$ and is close to zero at time lags $\tau>5$ seconds during
all four conditions. (Right) A post-stroke subject: $C(\tau)$ has a
much larger amplitude at $\tau=0$ which lasts for lags $\tau$ up to
20 seconds, indicating a strong coupling between the BP and BFV
signals, i.e., loss of cerebral autoregulation. } \label{fig03}
\end{figure*}

The BFV data show more random fluctuations for healthy subjects
compared to post-stroke subjects (one example is shown in
Fig.~\ref{fig01}), leading to slightly reduced cross-correlation
amplitude between BP and BFV for healthy subjects.  However, the
difference in the maximum cross-correlation value $C_{\rm max}$ is
not significant (see Table III), the reason being that the general
shape of the BP and BFV oscillations at each heartbeat is very
similar for both healthy and post-stroke subjects, and in addition,
the amplitude of BP and BFV oscillations is much larger relative to
the small random fluctuations on top of these oscillations. In
contrast, the shape of the oscillations observed in the phase
increments of the BP and BFV signals at every heartbeat is very
different for healthy subjects [see Figs.~\ref{fig02}(c),
\ref{fig02}(d)] but very similar for post-stroke subjects [see
Figs.~\ref{fig02}(g), \ref{fig02}(h)]. Thus, the instantaneous phase
cross-correlation between BP and BFV for healthy subjects is
significantly different from post-stroke subjects, as one can see by
comparing Fig.~\ref{fig01n} and Fig.~\ref{fig03}. This difference is
also shown in Table III. The comparative statistical tests indicate
significant difference between the control and stroke group based on
the instantaneous phase cross-correlation parameter $C_{\rm max}$
when we compare both the stroke side and the normal side of
post-stroke subjects with healthy subjects. There is no significant
difference in $C_{\rm max}$ obtained from the direct
cross-correlation between BP and BFV signals when we compare the
normal side of post-stroke subjects with healthy subjects (see also
results in Sec.~\ref{res_pha_corr}).

We also note that, healthy subjects exhibit higher variability in
the intervals between consecutive heartbeats, leading to higher
variability in the intervals between consecutive peaks in the BP and
BFV waves (Fig.~\ref{fig01}). As a result, both the direct
cross-correlation and the instantaneous phase cross-correlation
between the BP and BFV signals decay faster with increasing the time
lag $\tau$ for healthy subjects compared to post-stroke subjects. For
both methods we show that the parameter $\tau_0$, which
characterizes the decay in the cross-correlations, separates equally
well the stroke group from the control group (see Table III).

\subsection{Phase Synchronization Analysis}
\label{res_syn}

In Fig.~\ref{fig01n2} we present the results of the phase
synchronization analysis for four healthy and four post-stroke subjects.
The statistics for all four physiologic conditions (supine rest,
tilt, hyperventilation, CO$_2$ rebreathing) in our database and over
the entire control and stroke groups are summarized in Table III. We
find a statistically significant difference between the control and
stroke group based on the entropy index $\rho$.

In the phase synchronization method one estimates the difference
between the instantaneous phases of two signals {\it at the same
time}. Thus, the phase synchronization method does not reflect
possible time delays between the two studied signals, and
correspondingly, between their instantaneous phases. Such time
delays can often be observed in coupled physical and physiological
systems~\cite{Rybski03,Cimponeriu04}. To demonstrate the effect of
such time delay in context of the system we study, we consider the
difference between the instantaneous phases of BP and BFV, when the
phase of BFV is taken at time $t_1$ and the phase of BP is taken at
time $t_2=t_1+\tau$. As we show in Fig.~\ref{fig01n3}, the result of
the phase synchronization analysis is very different when
considering instantaneous phase difference with a time delay --- the
histogram of the phase difference becomes much broader (the entropy
index $\rho$ much smaller) compared to the case when no time delay
is introduced.

Statistics for the groups and physiologic conditions  are presented
in Table III, and indicate a very different result --- no
statistically significant difference between the control and stroke
group for the entropy index $\rho$ estimated for an arbitrary chosen
time delay of $\tau=0.4$ seconds. This is in contrast to the results
obtained for the entropy index $\rho$ from the synchronization
analysis when one does not consider a time delay. Thus, the results
of the traditional phase synchronization method strongly depend on
the time at which the instantaneous phases of two signals are
compared~\cite{Rybski03,Cimponeriu04}. This motivates our approach
to investigate the cross-correlation between the instantaneous
phases of two coupled systems at different time lag $\tau$, as
presented in Sec.~\ref{res_pha_corr}.

\begin{table*}
{
\scriptsize
\begin{tabular}{c|c|cc|cc|cc|cc|cc}
\hline Variable & &\multicolumn{2}{c|}{Supine}  &
\multicolumn{2}{c|}{Tilt} & \multicolumn{2}{c|}{Hyperventilation}
&\multicolumn{2}{c|}{CO$_2$ rebreathing} &
\multicolumn{2}{c}{Statistics}\\\hline

& & BFV- & BFV- & BFV-&BFV- & BFV- & BFV-& BFV- & BFV-&
BFV- & BFV-\\

& & MCAR/ & MCAL/ &MCAR/ &MCAL/ &MCAR/ &MCAL/ &MCAR/ &MCAL/ &MCAR/
&MCAL/\\

& & Normal & Stroke &Normal &Stroke &Normal &Stroke &Normal
&Stroke &Normal &Stroke\\

& & side & side & side & side & side & side & side & side & side &
side\\\hline




 direct cross-correl.& Control & $0.91\pm0.02$ & $0.89\pm0.06$ & $0.88\pm0.05$ & $0.84\pm0.06$& $0.88\pm0.13$& $0.87\pm0.12$& $0.91\pm0.03$& $0.91\pm0.03$& $0.066^{*}$& $0.069^{*}$\\
$C_{\rm max}$& Stroke& $0.92\pm0.03$ & $0.92\pm0.03$ & $0.88\pm0.09$ & $0.89\pm0.06$& $0.91\pm0.06$& $0.90\pm0.09$& $0.92\pm0.04$& $0.93\pm0.04$& $0.33^{\dagger}$&$0.024^{\dagger}$ \\\hline

 direct cross-correl.& Control & $6.5\pm4.1$  & $6.6\pm4.1$ & $3.3\pm1.4$ & $3.5\pm1.3$& $3.2\pm1.2$& $3.2\pm1.3$& $4.8\pm3.2$& $5.0\pm3.4$& $0.008^{*}$& $0.01^{*}$\\
 $\tau_0$& Stroke & $11.4\pm6.6$ & $11.2\pm6.9$& $6.5\pm5.6$ & $6.4\pm5.6$& $6.2\pm5.8$& $6.2\pm5.7$& $9.8\pm7.2$& $9.6\pm7.2$& $0.0001^{\dagger}$&$0.0004^{\dagger}$ \\\hline


phase synch.& Control & $0.32\pm0.07$ & $0.31\pm0.06$ & $0.25\pm0.07$ & $0.23\pm0.07$& $0.21\pm0.08$& $0.20\pm0.08$& $0.27\pm0.06$& $0.26\pm0.06$& $<0.0001^{*}$& $0.0002^{*}$\\
index $\rho$ for $\tau=0$ &Stroke & $0.35\pm0.07$ & $0.34\pm0.07$ & $0.26\pm0.09$ & $0.27\pm0.09$& $0.25\pm0.09$& $0.25\pm0.10$& $0.32\pm0.09$& $0.32\pm0.09$& $0.047^{\dagger}$&$0.008^{\dagger}$ \\\hline

phase synch.& Control & $0.18\pm0.05$ & $0.17\pm0.05$ & $0.11\pm0.05$ & $0.11\pm0.06$& $0.11\pm0.06$& $0.11\pm0.06$& $0.12\pm0.05$& $0.12\pm0.04$& $<0.0001^{*}$& $<0.0001^{*}$\\
index $\rho$ for $\tau=0.4$s& Stroke & $0.21\pm0.05$ & $0.21\pm0.05$ & $0.11\pm0.05$ & $0.12\pm0.06$& $0.11\pm0.06$& $0.11\pm0.06$& $0.13\pm0.05$& $0.13\pm0.05$& $0.43^{\dagger}$&$0.18^{\dagger}$ \\\hline

phase cross-correl.& Control & 0.55$\pm$0.17& 0.49$\pm$0.22& 0.39$\pm$0.16& $0.33\pm0.17$& $0.47\pm0.17$& $0.43\pm0.16$& $0.45\pm0.13$& $0.41\pm0.13$& 0.006$^{*}$& 0.07$^{*}$\\
$C_{\rm max}$          & Stroke  & 0.62$\pm$ 0.12& 0.58$\pm$0.14& 0.43$\pm$0.21& $0.47\pm0.20$& $0.51\pm0.19$& $0.52\pm0.20$& $0.57\pm0.20$&$0.58\pm0.21$& 0.049$^{\dagger}$&0.001$^{\dagger}$ \\\hline
phase cross-correl.& Control & $2.0\pm2.0$& $2.0\pm1.7$& $2.2\pm1.3$& $2.5\pm1.5$& $2.0\pm0.9$& $2.0\pm0.9$& $2.6\pm1.5$& $2.6\pm1.5$& 0.6$^{*}$& 0.7$^{*}$ \\
$\tau_0$         & Stroke  & $6.6\pm6.7$& $5.9\pm5.3$& $3.8\pm3.4$& $3.6\pm3.5$& $5.3\pm5.6$& $5.7\pm5.9$& $5.9\pm5.4$& $5.8\pm5.6$&0.0003$^{\dagger}$& 0.0004$^{\dagger}$ \\\hline

\end{tabular}
}

* P value between physiologic conditions comparisons

$\dagger$ P value between groups comparisons

\caption{The maximum correlation strength $C_{\rm max}$ and the
characteristic lag $\tau_0$ obtained from the direct
cross-correlation of the BP and BFV signals and from the
cross-correlation of the instantaneous phase increments in the BP
and BFV signals, as well as the entropy index $\rho$ obtained from
the synchronization analysis for the control and stroke group during
different physiologic conditions. We compare data from BFV in the
right MCA (BFV-MCAR) in healthy subjects with data from BFV in the
normal side MCA in post-stroke subjects (mean $\pm$ standard
deviation values are presented in the left column for each
condition). We compare data from BFV in the left MCA (BFV-MCAL) in
healthy subjects with data from BFV in the stroke side MCA in
post-stroke subjects (mean $\pm$ standard deviation values are
presented in the right column for each condition). The p values from
2x4 MANOVA are calculated for comparing groups and conditions
difference. } \label{table3}
\end{table*}

\subsection{Cross-correlation of Instantaneous Phase Increments}
\label{res_pha_corr}

We apply the instantaneous phase increments cross-correlation
analysis to all four conditions and both study groups. We find that
the patterns of the cross-correlation function $C(\tau)$ of the
instantaneous phase increments $\Delta \varphi$ of the BP and BFV
signals are very different for the stroke group compared to the
control group. In general, the cross-correlation function $C(\tau)$
for the control group is characterized by smaller amplitude and
faster decay (time lag $\tau$ less than 10 seconds)
[Fig.~\ref{fig03}]. In contrast, for post-stroke subjects, the
amplitude of the cross-correlation function $C(\tau)$ is much larger
and decays much slower (over time lags larger than 10 seconds)
[Fig.~\ref{fig03}].
Thus for those post-stroke subjects the changes of the phase of BFV
will change in the approximately same way with that of BP signals,
indicating a strong synchronization.

The correlations at short time scales (less than 10 seconds) may be
partially attributed to the effect of heart rate ($\sim$1 sec) and
respiration ($\sim$5 seconds)--- i.e., they reflect the effect of
other body regulations (similar to``background noise'') on both BP
and BFV signals. When cerebral autoregulation is effective, because
of its fast-acting mechanism~\cite{Panerai98}, it may also
contribute to the significantly weaker cross-correlations at short
time scales we find in healthy subjects compared to post-stroke
subjects (Fig.~\ref{fig03}). However, the correlation due to the
above mechanisms will decrease very fast when increasing the time
lag $\tau$ between BP and BFV signals. Thus, the correlations at
long time scales ($>$10 seconds) observed for post-stroke subjects
(Fig.~\ref{fig03}), cannot be attributed to the effect of cerebral
autoregulation. Instead, the existence of such strong and sustained
cross-correlations may imply that BFV will passively follow the
changes of BP, thus indicating absence of vascular dilation or
constriction and impaired cerebral autoregulation for the
post-stroke subjects.

To quantitatively distinguish the control group from the stroke
group, we study the characteristics of the correlation function
$C(\tau)$ for all subjects. For each correlation function $C(\tau)$,
we first find $C_{\rm max}$, the maximal value of $C(\tau)$, which
tells the strength of the correlation. Then we choose a threshold
value, e.g., $r=0.3$, search for the maximum of $|C(\tau)|$ during
each heart beat period along both positive and negative lags, and
find two points in those maxima with largest time lags at which the
correlation are still above $rC_{\rm max}$. The average of absolute
values of these two points gives the characteristic time lags
$\tau_0$. From $\tau_0$ and $C_{\rm max}$ for all subjects and
during all physiologic conditions, one can confirm that the stroke
group tends to have larger $C_{\rm max}$ and longer time lag
$\tau_0$ compared to those for the control group.

We apply MANOVA to demonstrate whether $\tau_0$ and $C_{\rm max}$
are different for healthy and post-stroke subjects. The results are
shown in Table~\ref{table3}. We find that during tilt and
hyperventilation, the control and stroke groups are not
significantly different (p values $>0.05$). In contrast, during
supine rest and CO$_2$ re-breathing, the difference between the
control group and the stroke group becomes significant (p values
$<0.05$).


To explain the above difference in $\tau_0$ and $C_{\rm max}$ during
supine rest, we note that post-stroke subjects exhibit higher BP and
CO$_2$ mean values in baseline (see Table~\ref{table1}). Therefore,
the BP-BFV autoregulatory curve for post-stroke subjects is shifted to
the right, to higher BP values, while at the same time the plateau of
this curve is narrowed due to the higher level of
CO$_2$~\cite{Panerai98}. CO$_2$ rebreathing increases the level of the
CO$_2$ after the period of hyperventilation (hypocapnia), thus further
testing the reactivity of the CA which shows impaired vasodilatatory
responses in post-stroke subjects. In contrast, vasoconstrictor
responses to tilt-up hyperventilation are preserved.


\section{Discussion and Conclusions}\label{Conclusion}

In this study we investigate dynamics of cerebral autoregulation
from the relationship between the peripheral BP in the finger and
the cerebral BFV for a group of healthy and post-stroke subjects during
the four different physiologic conditions of supine, tilt,
hyperventilation, CO$_2$ rebreathing. The mechanism of cerebral
autoregulation is traditionally accessed through the response of the
average cerebral BFV to abrupt perturbation in the the average BP
(e.g., upright tilt from supine position), and is typically
characterized by the ability of the cerebral blood vessels to
restore equilibrium and a steady cerebral blood flow after such
perturbation --- a process which is known to operate on time scales
above several heartbeats. In contrast, we focus on the dynamical
characteristics of the pressure-flow fluctuation around average
values. We show that in healthy subjects the CA mechanism is active
even during the steady equilibrium state. Moreover, we find that on
top of the BFV waveforms associated each heartbeat there are robust
fluctuations which are reduced in the post-stroke subjects,
indicating an active cerebral vascular regulation in healthy
subjects on time scales within a single heartbeat even under steady
BP.

To test for the dynamical patterns in the BP-BFV fluctuations we use
four different methods, and we compare the results of the analyses
by evaluating the combined effects of pressure autoregulation
(upright tilt) and metabolic autoregulation (hyperventilation and
CO$_2$ rebreathing) in healthy and post-stroke subjects. We find
that the gain and phase obtained from the traditional transfer
function analysis do not provide a significant difference between
healthy and post-stroke subjects. In contrast, the coherence is
significantly different in the heartbeat frequency range (0.7-1.4
Hz) when we compare both the normal side and the stroke side of
post-stroke subjects with healthy subjects.

Further, we find that the amplitude of the direct cross-correlation
between the BP and BFV signals does not separate the control from the
stroke group and between different physiologic conditions when we
compare the normal side of post-stroke subjects with healthy
subjects. However, comparing the stroke side of post-stroke subjects
with healthy subjects, we find a statistically significant difference,
suggesting that direct cross-correlation method is sensitive to detect
abnormalities in CA only for the stroke side. In addition, we observe
a significantly faster decay in the BP-BFV cross-correlation function
for healthy subjects, reflecting higher beat-to-beat variability in
the BP and BFV signals, compared to post-stroke subjects.

Since both BP and BFV are driven by the heartbeat, we also test the
coupling between these two signals applying the phase
synchronization method. For healthy subjects we observe a weaker
synchronization between BP and BFV characterized by lower value of
the synchronization index $\rho$ compared to post-stroke subjects,
indicating that the CA mechanism modulates the cerebral BFV waves so
that they are not completely synchronized with peripheral BP waves
driven by the heartbeat. We find a significantly stronger BP-BFV
synchronization in post-stroke subjects associated with loss of
cerebral autoregulation.

To probe how the CA mechanism modulates the BFV we investigate the
dynamical patterns in the instantaneous phase increments of the BP
signals $\Delta \varphi_{BP}$ and compare them with the patterns we
find in the instantaneous phase increments of the BFV signals
$\Delta \varphi_{BFV}$.  Remarkably, we find that for post-stroke
subjects $\Delta \varphi_{BP}$ and $\Delta \varphi_{BFV}$ exhibit
practically identical patterns in time, leading to very high degree
of cross-correlation between $\Delta \varphi_{BP}$ and $\Delta
\varphi_{BFV}$. In contrast, for healthy subjects $\Delta
\varphi_{BFV}$ exhibits robust random fluctuations very different
from the structured oscillatory patterns we find in $\Delta
\varphi_{BP}$. This leads to a significantly reduced
cross-correlation between $\Delta \varphi_{BP}$ and $\Delta
\varphi_{BFV}$ for healthy subjects compared to post-stroke
subjects. Our results indicate a statistically significant
separation between the stroke and the control group when we compare
both the normal and the stroke side in post-stroke subjects with
healthy subjects. This suggests that our new approach based on
cross-correlation of the instantaneous phase increments of BP and
BFV is sensitive to detect impairment of cerebral vascular
autoregulation in both hemispheres for subjects with minor ischemic
stroke.

Our findings of robust fluctuations in $\Delta \varphi_{BFV}$, which
do not synchronize with periodic patterns in $\Delta \varphi_{BP}$,
clearly indicate that the mechanism of cerebral autoregulation impacts
the dynamics not only on scales above several heartbeats but is also
active within a single heartbeat, i.e., much shorter time scales than
previous known. Since these fluctuations are present in the data after
we have truncated the a part of BP and BFV signals corresponding to
the initial perturbation related to changes in the physiologic
condition, our results also suggest that the cerebral autoregulation
plays an important role even in the quasi-steady state.


\section*{Acknowledgments}
Z.C., K.H., H.E.S., and P.Ch.I. acknowledge support from NIH Grant
No. HL071972, No. 2RO1 HL071972 and NIH/National Center for Research
Resources Grant No. P41RR13622. V.N. acknowledges support from CIMIT -
New Concept grant (W81XWH), American Heart Foundation Grant No. 99
30119N, 1R01 NIH-NINDS (1R01-NS045745-01), NIH GCRC Grant 5 MOIRR00034
and MO1-RR01302, and The Older American Independence Center Grant 2P60
AG08812-11.






\end{document}